\documentclass[12pt]{article}
\usepackage{graphics,graphicx}
\usepackage{amssymb,epsfig,amsmath,euscript,array}
\usepackage{cite}
\usepackage{axodraw}
\usepackage{pstricks}
\usepackage{color}

\makeatletter
\@addtoreset{equation}{section}
\makeatother
\renewcommand{\theequation}{\thesection.\arabic{equation}}

\newcommand{\startappendix}{
\setcounter{section}{0}
\renewcommand{\thesection}{\Alph{section}}
\renewcommand{\theequation}{\Alph{section}.\arabic{equation}}}

\newcounter{multieqs}

\newcommand{\be}{\begin{equation}}
\newcommand{\ee}{\end{equation}}

\newcommand{\bm}[1]{\mbox{\boldmath $#1$}}

\newcommand{\kslash}{k \!\!\! / }

\newcommand{\lslash}{l \!\! / }
\newcommand{\Pslash}{P \!\!\!\! / }

\newcommand{\islash}{i \!\!\! / }
\newcommand{\jslash}{j \!\!\! / }
\newcommand{\aslash}{a \!\!\! / }
\newcommand{\bslash}{{b \hspace{-6pt} \slash} }

\newcommand{\onslash}{1 \!\!\! / }
\newcommand{\twslash}{2 \!\!\!/ }
\newcommand{\thslash}{3 \!\!\!/ }
\newcommand{\foslash}{4 \!\!\! / }
\newcommand{\fislash}{5 \!\!\! / }

\newcommand{\mslash}{m \!\!\! / }

\def\bd{\begin{document}}
\def\ed{\end{document}}
\def\nn{\nonumber}
\def\bea{\begin{eqnarray}}
\def\eea{\end{eqnarray}}

\def\red{\color{red}}
\def\black{\color{black}}
\def\blue{\color{blue}}
\def\orange{\color{orange}}

\def\ab{(ijab)}
\def\ba{(ijba)}
\def\ijab{{\tr}_{+}(\islash\, \jslash\, \aslash \, \bslash)}
\def\ijba{{\tr}_{+}(\islash\, \jslash\, \bslash \, \aslash)}
\def\ijaP{{\tr}_{+}(\islash\, \jslash\, \aslash \, \Pslash)}
\def\ijPLa{{\tr}_{+}(\islash\, \jslash\, \Pslash_L \, \aslash)}
\def\ijaPL{{\tr}_{+}(\islash\, \jslash\, \aslash \, \Pslash_L)}
\def\ijPLza{{\tr}_{+}(\islash\, \jslash\, \Pslash_{L;z} \, \aslash)}
\def\ijaPLz{{\tr}_{+}(\islash\, \jslash\, \aslash \, \Pslash_{L;z})}
\def\ijPa{{\tr}_{+}(\islash\, \jslash\, \Pslash \, \aslash)}
\def\iaPb{{\tr}_{+}(\islash\, \aslash\, \Pslash \, \bslash)}
\def\ibPa{{\tr}_{+}(\islash\, \bslash\, \Pslash \, \aslash)}
\def\ijPmu{{\tr}_{+}(\islash\, \jslash\, \Pslash \, \mu)}
\def\ibmuP{{\tr}_{+}(\islash\, \bslash\, \mu \, \Pslash)}
\def\ibmua{{\tr}_{+}(\islash\, \bslash\, \mu \, \aslash)}
\def\iamub{{\tr}_{+}(\islash\, \aslash\, \mu \, \bslash)}
\def\jaPb{{\tr}_{+}(\jslash\, \aslash\, \Pslash \, \bslash)}
\def\ijmuP{{\tr}_{+}(\islash\, \jslash\, \mu \, \Pslash)}
\def\ijmum{{\tr}_{+}(\islash\, \jslash\, \mu \, \mslash)}
\def\ijmmu{{\tr}_{+}(\islash\, \jslash\, \mslash \, \mu)}
\def\ijmP{{\tr}_{+}(\islash\, \jslash\, \mslash \, \Pslash)}
\def\iabP{{\tr}_{+}(\islash\, \aslash\, \bslash \, \Pslash)}
\def\ijbP{{\tr}_{+}(\islash\, \jslash\, \bslash \, \Pslash)}
\def\jbPa{{\tr}_{+}(\jslash\, \bslash\, \Pslash \, \aslash)}
\def\ijPb{{\tr}_{+}(\islash\, \jslash\, \Pslash \, \bslash)}
\def\jbmua{{\tr}_{+}(\jslash\, \bslash\, \mu \, \aslash)}

\def\loablt{ {\tr}_{+}(\lslash_1\, \aslash \, \bslash\, \lslash_2)}

\def\ijlolt{{\tr}_{+}(\islash\, \jslash\, \lslash_1 \, \lslash_2)}
\def\ijltlo{{\tr}_{+}(\islash\, \jslash\, \lslash_2 \, \lslash_1)}
\def\ibloa{{\tr}_{+}(\islash\, \bslash\, \lslash_1 \, \aslash)}
\def\jaltb{{\tr}_{+}(\jslash\, \aslash\, \lslash_2 \, \bslash)}
\def\ialtb{{\tr}_{+}(\islash\, \aslash\, \lslash_2 \, \bslash)}
\def\bltloa{{\tr}_{+}(\bslash\, \lslash_2\, \lslash_1 \, \aslash)}
\def\jbloa{{\tr}_{+}(\jslash\, \bslash\, \lslash_1 \, \aslash)}
\def\ibPb{{\tr}_{+}(\islash\, \bslash\, \Pslash \, \bslash)}
\def\ijltb{{\tr}_{+}(\islash\, \jslash\, \lslash_2 \, \bslash)}

\def\ijloa{{\tr}_{+}(\islash\, \jslash\,  \lslash_1 \, \aslash)}
\def\ijblt{{\tr}_{+}(\islash\, \jslash\,  \bslash \, \lslash_2)}

\def\jakb{{\tr}_{+}(\jslash\, \aslash\, \kslash \, \bslash)}
\def\iakb{{\tr}_{+}(\islash\, \aslash\, \kslash \, \bslash)}

\def\tofo{{\tr}_{+}(\onslash\, \thslash\, \twslash \, \foslash)}
\def\foto{{\tr}_{+}(\onslash\, \thslash\, \foslash \, \twslash)}
\def\tofi{{\tr}_{+}(\onslash\, \thslash\, \twslash \, \fislash)}
\def\fito{{\tr}_{+}(\onslash\, \thslash\, \fislash \, \twslash)}

\def\lrangle#1#2{\langle #1\,#2\rangle}

\def\Li{{$\rm Li}_2$}
\def\eps{\epsilon}
\def\epsuv{{\epsilon_{\rm \mbox{\tiny UV}}}}
\let\bm=\bibitem
\let\la=\label

\def\npb#1#2#3{Nucl. Phys. {\bf{B#1}} #3 (#2)}
\def\plb#1#2#3{Phys. Lett. {\bf{#1B}} #3 (#2)}
\def\prl#1#2#3{Phys. Rev. Lett. {\bf{#1}} #3 (#2)}
\def\prd#1#2#3{Phys. Rev. {D \bf{#1}} #3 (#2)}
\def\cmp#1#2#3{Comm. Math. Phys. {\bf{#1}} #3 (#2)}
\def\cqg#1#2#3{Class. Quantum Grav. {\bf{#1}} #3 (#2)}
\def\nppsa#1#2#3{Nucl. Phys. B (Proc. Suppl.) {\bf{#1A}}#3 (#2)}
\def\ap#1#2#3{Ann. of Phys. {\bf{#1}} #3 (#2)}
\def\ijmp#1#2#3{Int. J. Mod. Phys. {\bf{A#1}} #3 (#2)}
\def\rmp#1#2#3{Rev. Mod. Phys. {\bf{#1}} #3 (#2)}
\def\mpla#1#2#3{Mod. Phys. Lett. {\bf A#1} #3 (#2)}
\def\jhep#1#2#3{J. High Energy Phys. {\bf #1} #3 (#2)}
\def\atmp#1#2#3{Adv. Theor. Math. Phys. {\bf #1} #3 (#2)}
\newcommand{\EQ}[1]{\begin{equation} #1 \end{equation}}
\newcommand{\AL}[1]{\begin{subequations}\begin{align} #1 \end{align}\end{subequations}}
\newcommand{\SP}[1]{\begin{equation}\begin{split} #1 \end{split}\end{equation}}
\newcommand{\ALAT}[2]{\begin{subequations}\begin{alignat}{#1} #2 \end{alignat}
                        \end{subequations}}
\def\beqa{\begin{eqnarray}}
\def\eeqa{\end{eqnarray}}
\def\beq{\begin{equation}}
\def\eeq{\end{equation}}
\def\sst{\scriptscriptstyle}
\def\thetabar{\bar\theta}
\def\Tr{{\rm Tr}}
\def\one{\mbox{1 \kern-.59em {\rm l}}}
 \def\Nh{\hat{N}}

\newcommand{\half}{{\textstyle {1 \over 2}}}

\def\a{\alpha}      \def\da{{\dot\alpha}}
\def\b{\beta}       \def\db{{\dot\beta}}
\def\c{\gamma}  \def\G{\Gamma}  \def\cdt{\dot\gamma}
\def\d{\delta}  \def\D{\Delta}  \def\ddt{\dot\delta}
\def\e{\epsilon}        \def\vare{\varepsilon}
\def\f{\phi}    \def\F{\Phi}    \def\vvf{\f}
\def\h{\eta}
\def\k{\kappa}
\def\l{\lambda} \def\L{\Lambda}
\def\m{\mu} \def\n{\nu}
\def\o{\omega}
\def\p{\pi} \def\P{\Pi}
\def\r{\rho}
\def\s{\sigma}  \def\S{\Sigma}
\def\t{\tau}
\def\th{\theta} \def\Th{\Theta} \def\vth{\vartheta}
\def\X{\Xeta}
\def\z{\zeta}
\def\de{\partial}

\def\cA{{\cal A}} \def\cB{{\cal B}} \def\cC{{\cal C}}
\def\cD{{\cal D}} \def\cE{{\cal E}} \def\cF{{\cal F}}
\def\cG{{\cal G}} \def\cH{{\cal H}} \def\cI{{\cal I}}
\def\cJ{{\cal J}} \def\cK{{\cal K}} \def\cL{{\cal L}}
\def\cM{{\cal M}} \def\cN{{\cal N}} \def\cO{{\cal O}}
\def\cP{{\cal P}} \def\cQ{{\cal Q}} \def\cR{{\cal R}}
\def\cS{{\cal S}} \def\cT{{\cal T}} \def\cU{{\cal U}}
\def\cV{{\cal V}} \def\cW{{\cal W}} \def\cX{{\cal X}}
\def\cY{{\cal Y}} \def\cZ{{\cal Z}}

\def\ua{\underline{\alpha}}
\def\ub{\underline{\phantom{\alpha}}\!\!\!\beta}
\def\uc{\underline{\phantom{\alpha}}\!\!\!\gamma}
\def\um{\underline{\mu}}
\def\ud{\underline\delta}
\def\ue{\underline\epsilon}
\def\una{\underline a}\def\unA{\underline A}
\def\unb{\underline b}\def\unB{\underline B}
\def\unc{\underline c}\def\unC{\underline C}
\def\und{\underline d}\def\unD{\underline D}
\def\une{\underline e}\def\unE{\underline E}
\def\unf{\underline{\phantom{e}}\!\!\!\! f}\def\unF{\underline F}
\def\unm{\underline m}\def\unM{\underline M}
\def\unn{\underline n}\def\unN{\underline N}
\def\unp{\underline{\phantom{a}}\!\!\! p}\def\unP{\underline P}
\def\unq{\underline{\phantom{a}}\!\!\! q}
\def\unQ{\underline{\phantom{A}}\!\!\!\! Q}
\def\unH{\underline{H}}

\def\As {{A \hspace{-6.4pt} \slash}\;}
\def\bs {{b \hspace{-6.4pt} \slash}\;}
\def\Ds {{D \hspace{-6.4pt} \slash}\;}
\def\ds {{\del \hspace{-6.4pt} \slash}\;}
\def\ss {{\s \hspace{-6.4pt} \slash}\;}
\def\ks {{ k \hspace{-6.4pt} \slash}\;}
\def\ps {{p \hspace{-6.4pt} \slash}\;}
\def\pas {{{p_1} \hspace{-6.4pt} \slash}\;}
\def\pbs {{{p_2} \hspace{-6.4pt} \slash}\;}
\def\Ps {{P \hspace{-6.4pt} \slash}\;}
\def\Qs {{Q \hspace{-6.4pt} \slash}\;}

\def\Fh{\hat{F}}
\def\Vh{\hat{V}}
\def\Xh{\hat{X}}
\def\ah{\hat{a}}
\def\xh{\hat{x}}
\def\yh{\hat{y}}
\def\ph{\hat{p}}
\def\xih{\hat{\xi}}
\def\psit{\tilde{\psi}}
\def\Psit{\tilde{\Psi}}
\def\tht{\tilde{\th}}
\def\lt{\tilde{\lambda}}
\def\hl{\hat{\lambda}}
\def\hlt{\hat{\tilde{\lambda}}}
\def\llt{\tilde{l}}
\def\At{\tilde{A}}
\def\Qt{\tilde{Q}}
\def\Rt{\tilde{R}}
\def\Nt{\tilde{N}}

\def\at{\tilde{a}}
\def\st{\tilde{s}}
\def\ft{\tilde{f}}
\def\pt{\tilde{p}}
\def\qt{\tilde{q}}
\def\vt{\tilde{v}}
\def\nt{\tilde{n}}

\def\delb{\bar{\partial}}
\def\bz{\bar{z}}
\def\bD{\bar{D}}
\def\bB{\bar{B}}

\def\bk{{\bf k}}
\def\bl{{\bf l}}
\def\bp{{\bf p}}
\def\bq{{\bf q}}
\def\br{{\bf r}}
\def\bx{{\bf x}}
\def\by{{\bf y}}
\def\bR{{\bf R}}
\def\bV{{\bf V}}

\def\d{\delta}\def\D{\Delta}\def\ddt{\dot\delta}
\def\pa{\partial} \def\del{\partial}
\def\xx{\times}
\def\uno{\mbox{1 \kern-.59em {\rm l}}}
\def\trp{^{\top}}
\def\inv{^{-1}}
\def\dag{{^{\dagger}}}
\def\pr{^{\prime}}
\def\lan{\langle}
\def\ran{\rangle}
\def\rar{\rightarrow}
\def\lar{\leftarrow}
\def\lrar{\leftrightarrow}
\newcommand{\0}{\,\!}      
\def\one{1\!\!1\,\,}
\def\im{\imath}
\def\jm{\jmath}
\newcommand{\tr}{\mbox{tr}}
\newcommand{\slsh}[1]{/ \!\!\!\! #1}
\def\vac{|0\rangle}
\def\lvac{\langle 0|}
\def\hlf{\frac{1}{2}}
\def\ove#1{\frac{1}{#1}}
\def\Box{\square}
\def\ZZ{\mathbb{Z}}
\def\CC#1{({\bf #1})}
\def\bcomment#1{}
\def\bfhat#1{{\bf \hat{#1}}}
\def\VEV#1{\left\langle #1\right\rangle}
\newcommand{\ex}[1]{{\rm e}^{#1}} \def\ii{{\rm i}}
\def\rr{{\rm r}} \def\rs{{\rm s}}\def\rv{{\rm v}}
\def\ri{{\rm i}}\def\rj{{\rm j}}
\newcommand{\lrbrk}[1]{\left(#1\right)}
\newcommand{\sfrac}[2]{{\textstyle\frac{#1}{#2}}}

\def\Li{{\rm Li}_2}

\font\mybb=msbm10 at 12pt
\def\bb#1{\hbox{\mybb#1}}

\font\myBB=msbm10 at 18pt
\def\BB#1{\hbox{\myBB#1}}

\begin{document}

\begin{flushright}
IPPP/09/03,
DCPT/09/06,
QMUL-PH-09-01
\end{flushright}

\vspace{3pt}

\begin{center}

{\Large \bf Two-Loop Polygon Wilson Loops  in $\mathcal{N}=4$ SYM  }

%
\vspace{11pt}
\end{center}

\hspace{-2.1cm}{\mbox {\bf C.~Anastasiou$^{a}$,  A.~Brandhuber$^{b}$, P.~Heslop$^{b}$,
V.~V.~Khoze$^{c}$,  B.~Spence$^{b}$ and G.~Travaglini$^{b}$}}%
\footnote{
{\tt babis@phys.ethz.ch, valya.khoze@durham.ac.uk, \{ \tt \!\!\!a.brandhuber, p.j.heslop,  w.j.spence, g.travaglini\}@qmul.ac.uk }
}

\begin{center}
{\small \em
\begin{itemize}
\item[\ \ \ \ \ \ $^a$]
Institute for Theoretical Physics, ETH Z\"{u}rich \\
8093 Z\"{u}rich, Switzerland
\item[\ \ \ \ \ \ $^b$]
Centre for Research in String Theory\\
Department of Physics,
Queen Mary, University of London\\
Mile End Road, London, E1 4NS,
United Kingdom\\
\item[\ \ \ \ \ \ $^c$]
Institute for Particle Physics Phenomenology,  Department of Physics,  \\
Durham University,
Durham, DH1 3LE, United Kingdom

\end{itemize}
}

\vspace{-8pt}

\vspace{23pt} {\bf Abstract}
\end{center}

\noindent
We compute for  the first  time
the two-loop corrections to arbitrary $n$-gon lightlike Wilson loops
in ${\cal N}=4$ supersymmetric Yang-Mills theory, using efficient  numerical
methods.
The calculation is motivated by the remarkable agreement between
the finite part of planar six-point MHV amplitudes and hexagon Wilson
loops which has been observed at two loops.
At $n=6$ we confirm   that the ABDK/BDS ansatz
must be corrected by adding a remainder function, which depends only
on conformally invariant  ratios of kinematic variables.
We numerically  compute remainder functions for $n=7,8$ and verify
dual conformal invariance.
Furthermore, we study  simple and multiple collinear limits of the Wilson loop remainder functions
and demonstrate that they have precisely the form required by the collinear factorisation
of the corresponding two-loop $n$-point amplitudes.
The number of distinct diagram topologies  contributing to the $n$-gon Wilson loops
does not increase with $n$, and there is a fixed number of  ``master integrals", which we have computed.
Thus we have essentially computed general polygon
Wilson loops, and if the correspondence with amplitudes continues to hold,
all planar $n$-point two-loop MHV amplitudes in the ${\cal N}=4$ theory.

\setcounter{page}{0}
\thispagestyle{empty}
\newpage


\section{Introduction }
\setcounter{footnote}{0}

A surprising  feature encountered
in the study of supersymmetric gauge theories is the existence of
an intriguing iterative structure in the higher-loop expansion of the Maximally Helicity
Violating (MHV) scattering amplitudes in
planar $\mathcal{N}=4$ super Yang-Mills (SYM) theory.
This iterative structure  was first discovered by Bern, Dixon, Kosower and one of the present authors (ABDK)
studying  collinear limits of maximally supersymmetric gauge theory amplitudes,   and in the
planar four-point MHV amplitude at two loops \cite{abdk}.
In the same paper,  it was also conjectured that the same  iterative structure should hold for two-loop MHV amplitudes
with an arbitrary number of external legs.

In a subsequent important development,
Bern, Dixon and Smirnov (BDS) proposed an all-loop resummed formula for the $n$-point MHV amplitude, which they were able
to confirm in an impressive three-loop calculation of the four-point amplitude \cite{bds}.
According to this conjecture, multi-loop amplitudes can be re-expressed
in terms of the one-loop amplitude and four kinematic-independent functions of the 't Hooft coupling.
One of these functions is the cusp anomalous dimension \cite{Polyakov:1980ca,Brandt:1981kf,Korchemsky:1985xj},
 for which  an all-order expression has been proposed in
\cite{Beisert:2006ez}.

The ABDK/BDS conjecture was further investigated in several papers.
In particular, it was confirmed in a two-loop calculation for the five-point amplitude  in \cite{2l5pt} --
a result which  is particularly non-trivial  since it implies a cancellation of certain parity-odd terms
in  the two-loop term of the logarithm of the amplitude.%
\footnote{The iteration for the parity-even terms had been proved earlier in \cite{Cachazo:2006tj}.}
It was also pointed out in \cite{Khoze:2005nd} that the amplitudes of the $\beta$-deformed $\cN=4$ theory with
real $\beta$ are  identical to those of the undeformed theory (modulo an irrelevant, overall phase), and as such
they will satisfy the ABDK/BDS iterative structure if  the corresponding undeformed amplitudes do.
Explicit expressions of the four-point amplitudes at four and five loops were
also derived in \cite{4l} and \cite{5l}, respectively, and, for the four-dimensional cut-constructible part of the
five-point amplitude at three loops in
\cite{svw}; these expressions will allow for further  tests
of the  BDS ansatz at four and five loops once the relevant integral functions have been evaluated
to the necessary degree of accuracy in $\epsilon$.

One of the key aspects of the ABDK/BDS conjecture is the appearance of the exponentiation of the one-loop result in the
complete perturbative answer.
In a remarkable paper \cite{am},  Alday and Maldacena succeeded in using   the  AdS/CFT correspondence to
provide a string theory formalism to address scattering amplitudes at strong coupling. In particular, their  calculation of
the four-point amplitude reproduced the strong-coupling limit of the BDS ansatz. It also provided a
string theory explanation for why planar scattering amplitudes at strong coupling exponentiate, through a semiclassical
calculation.
It was argued in \cite{Abel:2007mw} that the same exponentiation of \cite{am} should hold not only for  MHV amplitudes
but also for  non-MHV amplitudes, since
the helicity dependence of the amplitudes in the prefactor is unlikely to modify the semiclassical exponent in the path integral.
However, for the non-MHV case
the exponentiation can only occur at strong coupling, and is not apparent in perturbation theory.

The result of \cite{am} suggested that the vacuum expectation value of a  polygonal $n$-edged Wilson loop,
evaluated  this time at weak coupling, could  be related  to the perturbative $n$-point MHV amplitude in
$\mathcal{N}=4$ SYM \cite{dks,bht}. This was confirmed in a one-loop calculation for $n=4$ in \cite{dks} and subsequently
for arbitrary $n$ in \cite{bht}. Drummond, Henn, Korchemsky and Sokatchev (DHKS)  were later able to confirm this
conjecture in a remarkable analytic calculation of the two-loop four-edged Wilson loop \cite{dhks4}, followed by a
semi-analytical calculation of the five-edged Wilson loop \cite{dhks5}.

It was later argued \cite{Alday:2007he}   that the BDS ansatz may be incomplete, specifically for $n$-point amplitudes with $n\ge 6$
\cite{Bartels:2008ce}.
The authors of \cite{seven} have carried out an explicit calculation which shows that the BDS ansatz indeed needs to be
modified in order to reproduce the two-loop term of the logarithm of the six-point amplitude.
In a parallel development, the corresponding six-point lightlike Wilson loop was computed at two loops
in \cite{dhksbum,dhks6},  and compared  in \cite{seven, dhks6}
to  the parity-even part of the amplitude evaluated in \cite{seven}.%
\footnote{The full six-point amplitude at two loops has been presented in \cite{Cachazo:2008hp}.}
The result  of this analysis is that
the MHV amplitude, stripped of the tree-level prefactor, and the Wilson loop are in perfect agreement (up to an additive constant)
for the two-loop, six-point case,%
\footnote{More accurately, there is a difference in the coefficients of the subleading $1 / \e$ pole  for the Wilson loop and the
amplitude. We will come back to this point in Section \ref{WLremainderfunction}.  }
but there is an additional contribution compared to what the BDS ansatz predicts.
This extra term, which we will refer to as the remainder  function,
will be  one of the main characters of  our paper.

The possibility of having a nonvanishing remainder function was neatly explained  in \cite{dhks5} in terms of the
anomalous conformal symmetry of the lightlike Wilson loop.
In that paper the associated anomalous conformal Ward identities were derived, and it was shown that
the BDS expression provided a particular solution to these Ward identities.
At the same time, the
anomalous Ward identities cannot uniquely determine terms that are
invariant under the conformal symmetry, and this leaves room for a
conformally invariant remainder function \cite{dhks5}. For $n \leq
5$,  the lightlike constraints on the particle momenta restrict
such conformally invariant contributions to just
(kinematic-independent) constants. However, starting from $n=6$
edges one can build functions of the conformally invariant ratios
which are left  undetermined by the Ward identities, and need no
longer vanish. DHKS made the prediction therefore that, if the
duality with Wilson loops holds, the remainder function should
depend on the kinematics only through cross-ratios.

The dual conformal symmetry of the Wilson loop was also instrumental in suggesting that the $S$-matrix of the $\cN=4$ theory
should possess a dual superconformal symmetry \cite{dhks,dhksgen}, which is expected to be exact at tree level, and violated by an anomaly at the loop level. Indeed, it was later proved in \cite{bhtrec} using a supersymmetric version
\cite{ah,bhtrec,cahk} of the BCF recursion relation
\cite{bcf,bcfw} that  the tree-level  $S$-matrix of the planar $\cN=4$ theory is covariant under the dual superconformal symmetry.
A solution of the supersymmetric recursion relation of \cite{ah,bhtrec,cahk} was also  presented in \cite{drhe}.
We also mention that the dual superconformal charges resurface as part of an infinite tower of
charges coming from integrability of the dual AdS sigma model  \cite{bermal, tse}.

In this paper, we present the results of our study of the Wilson loop remainder function for arbitrary $n$.
One important observation is that the structure of infrared and other integrable singularities of the diagrams which enter the Wilson loop calculation does not change for $n>7$.
Therefore, with the same numerical routines we can evaluate Wilson loops for
arbitrary $n$. We should note that our calculations were performed for Euclidean kinematics,
but a generalisation to Minkowskian kinematics is possible.

There are several interesting properties of the remainder functions which we have analysed.
The first one  is its  conjectured  dependence on the kinematics only through cross-ratios. In our study we have collected ample numerical evidence that confirms this expectation for $n=6,7$ and $8$ points.
The second aspect is the study of simple and multiple collinear limits of this function.
Specifically, using universal factorisation theorems \cite{bddk,fusing,Bern:1995ix,Kosower:1999xi}
for scattering amplitudes one can predict \cite{seven} the behaviour of the $n$-point amplitude remainder functions under simple collinear limits, namely $\cR_n \to \cR_{n-1}$.
In this paper we will show that the appropriately defined Wilson loop remainder function $\cR_n^\mathrm{WL}$
has exactly the same collinear behaviour, namely
\beq
\label{cfowl}
\cR_n^\mathrm{WL} \, \to \, \cR_{n-1}^\mathrm{WL}
\ .
\eeq
Notice that no additional constant term appears on the right hand side of \eqref{cfowl}. We have
checked this numerically for $n=6,7$ and $8$ sided polygon Wilson loops.
Finally, one of the most important goals for the future is to find analytic expressions for the remainder
functions. As a first step we have initiated a detailed map of these functions, in particular for $n=6$, for
a wide range of values of the cross-ratios. We were able to make intriguing
observations for special values of the cross-ratios and lower dimensional slices of the kinematic
parameter spaces. On general grounds,  we expect the remainder functions to be transcendentality four functions of the
conformal cross-ratios. However, even if we restrict the remainder functions to a one-dimensional slice of the parameter space, the space of transcendentality four functions is rather large and, hence, numerical methods are not sufficient to determine the remainder function.
Clearly, new theoretical ideas, possibly from the AdS/CFT correspondence or integrability,  are needed in order
to make progress in this direction.

The rest of the paper is organised as follows. In Section 2 we review salient features of planar
gluon scattering amplitudes in $\cN=4$ SYM, their recursive properties at loop level, and
the BDS all-loop ansatz. Furthermore, we introduce the amplitude remainder function, which is the difference between the full amplitude and the BDS ansatz, and is expected to depend only on the dual conformal cross-ratios of
kinematic invariants. In Section 3 we set up the corresponding polygon Wilson loop calculations
at two loops,  and give a natural definition of the Wilson loop remainder function, which
behaves under collinear limits in the same way as the amplitude remainder function. In Section 4
we present details about the diagrams and the corresponding Feynman integrals entering the calculation of
arbitrary, lightlike, $n$-gon Wilson loops. In Section 5 we discuss the numerical evaluation of these Feynman integrals.
In Section 6 we present a detailed,
numerical analysis of the six-point remainder function including tests of dual conformal invariance, the
explicit values of the remainder function at specific values of the cross-ratios, and various illustrative plots.
In Section 7 we start off with a discussion of the seven-point remainder function, and give explicit
numerical results to illustrate our checks of dual conformal invariance and invariance under cyclic
permutations and reflections of the external momenta. We then move on to discuss simple and
multiple collinear limits of two-loop amplitudes and Wilson loops. Specifically, we present evidence that
in the simple collinear limits
the seven-point remainder function becomes equal to  the six-point remainder function.
Finally, in Section 8 we present a similar analysis  for the eight-point Wilson loop and briefly discuss the generalisation to arbitrary $n$.

\setcounter{footnote}{0}

\section{Planar amplitudes in $\cN=4$ super Yang-Mills and
the ABDK/BDS ansatz}

The infinite sequence of $n$-point
 planar
MHV amplitudes in
$\cN\!=\!4$    SYM  has a  remarkably simple form.
At any loop order $L$, the amplitude can be expressed as the tree-level amplitude, times a scalar, helicity-blind
function $\cM_n^{(L)}$:
\beq\label{fullampl}
\cA_{n}^{(L)} \ =  \cA^{\rm tree}_n\, \cM_n^{(L)}.
\eeq
At  one loop,  the function $\cM_n^{(1)}$
is  simply a sum  of  two-mass easy box functions $F^{\rm 2m\,e}$
\cite{Bern:1993kr}, with coefficient equal to one:
\beq
\label{Mfunction}
 \cM_{n}^{(1)}   \ = \  \sum_{p, q}
F^{\rm 2m\,e} (p, q, P, Q)
\ .
\eeq
In \cite{abdk}, ABDK discovered an intriguing  iterative structure
in the two-loop expansion of the MHV amplitudes at four points. This relation can be written as
\beq
\label{babis}
 \cM_4^{(2)}(\e)  - {1\over 2} \big( \cM_4^{(1)}(\e) \big)^2 \ = \ f^{(2)} (\e) \cM_4^{(1)} (2 \e ) + C^{(2)} + \cO (\e)
 \ ,
 \eeq
 where
 \beq
 \label{f2}
 f^{(2)} (\e) \ = \ -\zeta_2 - \zeta_3 \e - \zeta_4 \e^2
 \ ,
 \eeq and
 \beq
 \label{c2}
 C^{(2)}\  = \  -{1\over 2} \zeta_2^2
 \ .
 \eeq
In \cite{abdk}, it was  conjectured  that
\eqref{babis}-\eqref{c2}
should
hold for two-loop amplitudes with an arbitrary number of legs --
a conjecture which was  consistent with an
explicit evaluation of the universal two-loop splitting amplitude.

Building upon the iterative relation of \cite{abdk},  and the known
universal infrared behaviour
of gauge theory amplitudes \cite{ir1,ir2,ir3,Ivanov:1985np,Korchemsky:1988si,ir4,ir5,ir6,ir7,ir8},
BDS  proposed a resummed, exponentiated expression for the
scalar function $\cM_{n}$. In the same paper, this conjecture was checked  in a three-loop calculation in
the four-point case.  Specifically, the BDS conjecture is  expressed as \cite{bds}
\beq
\label{bds}
\cM_n \ := \ 1 + \sum_{L=1}^{\infty} a^L \cM_{n}^{(L)} (\epsilon )  \ =  \
\exp \Big[ \sum_{L=1}^{\infty} a^L  \Big( f^{(L)} (\epsilon) \cM_{n}^{(1)} ( L \epsilon )  + C^{(L)} + E_n^{(L)}(\epsilon )\Big)
\Big]
\ ,
\eeq
where $a$ is the loop-counting parameter. In the conventions of
\cite{bds}, this is defined as
$ a=[{g^2 N/ (8 \pi^2)}] (4\pi e^{-\gamma})^\eps$ .
Here $f^{(L)}(\epsilon )$ is a set of functions,
\beq \label{fleps}
f^{(L)}(\epsilon ) \, :=\, f_0^{(L)} + f_1^{(L)} \epsilon + f_2^{(L)} \epsilon^2  \ ,
\eeq
one at each loop order, which appear in the exponentiated all-loop expression
for the  infrared divergences in generic amplitudes in dimensional regularisation \cite{ir6} (and generalise the function $f^{(2)}$  in \eqref{babis}).
In particular, $f_0^{(L)} = \gamma_{K}^{(L)} / 4$, where $\gamma_{K}$ is the cusp anomalous dimension,
\beq
\label{defgamK}
\gamma_{K} (a) =\, \sum_{L=1}^{\infty} \, a^L\, \gamma_{K}^{(L)}\ , \qquad
\gamma_{K}^{(1)}=4 \ , \quad \gamma_{K}^{(2)}= -4\,  \zeta_2 \ ,
\eeq
related to the anomalous dimension of twist-two operators at large spin \cite{Korchemsky:1988si}.
The appearance of the cusp anomalous dimension characterises  the relation
between the infrared divergences of scattering amplitudes and ultraviolet divergences
of Wilson loops with cusps which was originally discussed in QCD in \cite{Korchemsky:1985xj,Ivanov:1985np}.

The $\cO ( \epsilon) $ term in \eqref{fleps} is related to the so-called collinear anomalous dimension $G$,  $f_1^{(L)}= (L/2)G^{(L)}$,
\beq
\label{defGcoll}
G (a) =\, \sum_{L=2}^{\infty} \, a^L\, G^{(L)}\ , \qquad
G^{(2)}= - \zeta_3 \ ,
\eeq
and  $f_2^{(2)}= -\zeta_4$. In  particular,  $f_2^{(2)}$ can already be  found  from simple collinear limits of the amplitudes.
For future reference  we also define
\beq
\label{defCa}
C (a) =\, \sum_{L=2}^{\infty} \, a^L\, C^{(L)}  \ .
\eeq
Importantly,  the constants $C^{(L)}$,  $f_0^{(L)}$, $f_1^{(L)}$ and $f_2^{(L)}$
on the right hand side of \eqref{bds} do not depend either on kinematics or on the number of particles $n$.
On the other hand, the non-iterating contributions  $E_n^{(L)}$ depend explicitly on $n$, but vanish as $\epsilon \to 0$.

BDS also suggested  a resummed expression for the appropriately defined finite part of the $n$-point MHV amplitude,
\beq
\label{finite}
\mathcal{F}_n \ =  e^{F^\mathrm{BDS}_n   }
\ ,
\eeq
where
\beq
\label{fbds}
F^\mathrm{BDS}_n (a) = \, {1\over 4} \gamma_K(a)\, \, F^{(1)}_n (0)  + C(a)
\ .
\eeq
The quantities $\gamma_K(a)$ and $C(a)$, are
given in  \eqref{defgamK} and  \eqref{defCa};
the entire dependence on kinematics of the BDS ansatz  enters through
the finite part of the one-loop box function,
$F^{(1)}_n (0)$.
Explicitly, one has \cite{bds}
\begin{align}\label{m1}
\cM_n^{(1)}(\epsilon)&= -{1\over 2 \e^2}\sum_{i=1}^n   \left( - {t_i^{[2]}
    \over \mu_\mathrm{}^2 }\right)^{-\e} +\ F_n^{(1)}(\epsilon)\ ,\\
F_n^{(1)}(0)& = {1 \over 2} \sum_{i=1}^n g_{n,i} \ ,
\end{align}
where
\beq
g_{n,i}  \ = \
-\sum_{r=2}^{[ n/2 ]  -1}
  \ln \left( { -t^{[r]}_{i}\over -t^{[r+1]}_{i} }\right)
  \ln \left({ -t^{[r]}_{i+1}\over -t^{[r+1]}_{i} }\right) \, + \,
D_{n,i} \, + \, L_{n,i} + {3\over 2} \zeta_2 \  ,
\eeq
and  $t^{[r]}_{i} := (p_i + \cdots + p_{i+r-1})^2$
are the kinematical invariants.  The explicit forms of the functions $D_{n,i}$ and $L_{n,i}$
depend on  whether $n$ is odd or even.
For $n=2m$ one has
\beqa
\label{LDeven}
D_{2m,i} &=& -\sum_{r=2}^{m-2}
\Li \left( 1- { t^{[r]}_{i} t^{[r+2]}_{i-1}
\over t^{[r+1]}_{i} t^{[r+1]}_{i-1} }  \right)
- {1 \over 2} \Li \left( 1- { t^{[m-1]}_{i} t^{[m+1]}_{i-1}
\over t^{[m]}_{i} t^{[m]}_{i-1}} \right) \ ,
\\ \nonumber
L_{2m,i} &=& - {1\over 4}
  \ln \left({ -t^{[m]}_{i}\over -t^{[m]}_{i+m+1}  } \right)
  \ln \left({ -t^{[m]}_{i+1}\over -t^{[m]}_{i+m} } \right) \ ,
\eeqa
whilst, when  $n=2m+1$,
\beqa
\label{LDodd}
D_{2m+1,i} &=& -\sum_{r=2}^{m-1}
\Li  \left( 1- { t^{[r]}_{i} t^{[r+2]}_{i-1}
\over t^{[r+1]}_{i} t^{[r+1]}_{i-1} } \right)  \ ,
\\ \nonumber
L_{2m+1,i} &=& -{ 1\over 2}
  \ln \left({ -t^{[m]}_{i}\over -t^{[m]}_{i+m+1}  } \right)
  \ln \left({ -t^{[m]}_{i+1}\over -t^{[m]}_{i+m} } \right)
  \ .
\eeqa
The case  $n=4$ is special; in this case the finite remainder is given by
\beq
F_4^{(1)}(0)
\ = \  {1\over 2} \ln^2\left({s\over t}\right) \, + \, 4 \zeta_2
\ ,
\eeq
which is the finite part of the zero-mass box function plus a constant shift.

\subsection{The amplitude remainder function and $n$-point cross-ratios} \label{sec:remainder-function}

Beyond five points, and starting from two loops,   the ABDK/BDS ansatz \eqref{babis}  needs to be modified by the addition of
a remainder function $\cR_n$ \cite{seven,dhks6},
\beq
\label{fm2}
\cM^{(2)}_n(\e ) - {1\over 2} \Big( \cM^{(1)}_n (\e) \Big)^2 \ = \
f^{(2)} (\e) \cM^{(1)}_n  ( 2 \e ) \, + \,  C^{(2)} \, + \, \cR_n \
\, + \, \cO (\e )
\ .
\eeq
We now move on to characterise it.

To begin with, we  recall
one of the important properties of the ansatz, namely that
it already incorporates the correct simple collinear limits of the
amplitude for all $n$ \cite{abdk,bds}.
We will review this in Section \ref{collinearsection}, but we would like to anticipate here one important consequence of this, namely the fact that
the remainder function must have trivial collinear limits.
With the definition given above of the
remainder function, one expects that under a simple collinear limit
\cite{seven}
\begin{equation}
\label{abbc}
  \cR_n \rightarrow \cR_{n-1} \ ,
\end{equation}
where no constant term can appear on the right hand side of
\eqref{abbc}.

An important advance was made in \cite{dhks5}, where it was realised that
the BDS ansatz is  a solution to the anomalous  Ward identity for the Wilson loop associated to the dual conformal symmetry.
As is by now common,
 one introduces dual (or region) momenta  $x_1, \cdots, x_n$  \cite{magic,am}
 and defines
the particles' momenta as
\beq
\label{dm}
p_i \ := \ x_i - x_{i+1}
\ ,
\eeq
(with the identification $x_1 = x_{n+1}$), which satisfy $n$ on-shell relations
$(x_i - x_{i+1})^2 = 0$, $i=1, \ldots, n$.
The dual conformal group then acts on the dual momenta \cite{dhks4, dhks5}.
It is important to notice that acting with conformal transformations on
the dual momenta $x$ does not endanger momentum conservation, which is
automatically satisfied once the momenta are written in the form \eqref{dm}.

Obviously, adding to the ABDK/BDS ansatz  any arbitrary function of  the
conformally invariant cross-ratios
\beq
{x_{ij}^2 x_{kl}^2 \over x_{ik}^2 x_{jl}^2}\ ,
\eeq
preserves conformal invariance and hence  would  give  another solution to the same conformal Ward identity.
With this in mind,  it was  argued
in \cite{dhks5} that the remainder function, if non-vanishing,
should depend on the kinematics of the scattering only through cross-ratios.
It is therefore important to examine how many independent  cross-ratios one can build at $n$ points.

Starting from  a set of $n$ arbitrary points, one can  define $n(n-3)/2$ cross-ratios.
This number is reduced to $n(n-5)/2$ if one imposes the $n$ on-shell conditions
for lightlike momenta -- notice that this is the same
as the number of two-mass easy box functions which could potentially appear in  a colour-ordered Yang-Mills
$n$-point amplitude.
If one considers four-dimensional external momenta we have
additional constraints that  the Gram determinant of any five  of them
should vanish. With the Gram determinant constraints, the number of possible on-shell cross-ratios is  reduced to \cite{dhks5}  $3n-15$  for $n>5$.
Incidentally, we note that the number of on-shell cross-ratios one can construct from $n$ points in a $D$-dimensional space with $D\geq n-1$ is also equal to $n(n-5)/2$ \cite{dhks5}. Discarding the Gram determinant constraints is therefore equivalent to
considering the external momenta as defined in a $D$-dimensional space with  $D\geq n-1$.

\begin{figure}[t]
\begin{center}
\scalebox{0.70}{
\includegraphics[width=12cm]{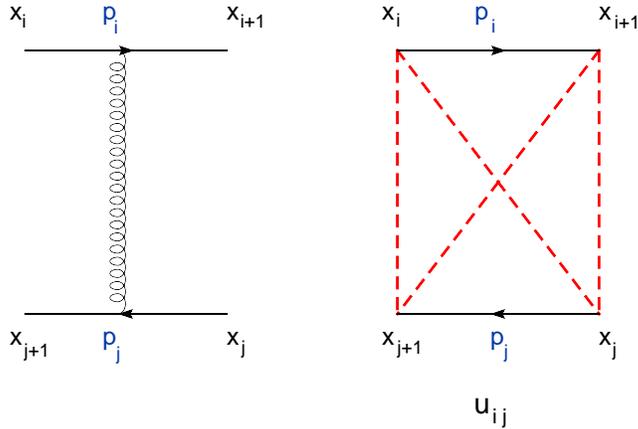}
}
\end{center}
\caption{\it On the left we represent the one-loop Wilson loop diagram which gives the
finite part of the two-mass easy box function with massless momenta $p_i,p_j$~{\rm \cite{bht}}.  On the right we represent the
  corresponding cross-ratio
  $u_{ij}$,  the red dashed lines depicting the factors $x_{ij}^2$,  $x_{i+1 j+1}^2$, $x_{i+1j}^2$, $x_{ij+1}^2$ in
  the definition of $u_{ij}$ in~\eqref{ourcrossratios}.}
\label{fig:cross-ratio}
\end{figure}

Taking this apparent coincidence
between the number of cross-ratios and  of two-mass easy boxes more seriously,
we define the independent  cross-ratios  we
will use to parametrise any $n$-point remainder function as
\beq
\label{ourcrossratios}
u_{i j} := {x_{i j+1}^2 x_{i+1 j}^2 \over x_{ij}^2 x_{i+1 j+1}^2}
\ .
\eeq
In~\cite{bht} it was shown explicitly how a finite one-loop Wilson loop diagram, where a gluon propagator
connects the momenta $p_i= x_i - x_{i+1} $ and $p_j= x_j - x_{j+1}$, reproduces the finite part of the
two-mass easy box function with kinematic invariants $s= x_{ij}^2$, $t= x_{i+1 j+1}^2$, $P^2 = x_{i j+1}^2$, $Q^2 = x_{i+1 j}^2$.
The choice of kinematic invariants appearing in \eqref{ourcrossratios} precisely matches those appearing in the
corresponding box, and  $u_{ij}$ is  of the form $ P^2 Q^2 / (s t )$
(see Figure~\ref{fig:cross-ratio})%
\footnote{Note also that a basis of
  off-shell cross-ratios can be given similarly, simply by allowing
  also the one-mass box   functions as well as the two-mass easy box
  functions.}.

Two comments are in order. Firstly, we observe that the ratios in \eqref{ourcrossratios} are the same as those entering
the  functions $D$ and $L$ defined in \eqref{LDeven}, \eqref{LDodd}, which appear
in the BDS ansatz  \eqref{fbds}.
Secondly, we mention that a generic cross-ratio, i.e.~one of the form
$x_{ij}^2 x_{lm}^2 / (x_{il}^2 x_{jm}^2)$ for generic $i, j, k, l$ ,
can be written as  a product of
cross-ratios of the form~\eqref{ourcrossratios}  for arbitrary $n$.
If we assume $i<l$
and $j<m$ this is
simply given by
\begin{equation}
  { x_{im}^2 x_{jl}^2\over  x_{ij}^2 x_{lm}^2} = \prod_{r=i}^{l-1}
  \prod_{s=j}^{m-1} u_{rs} \ .
\end{equation}

The two different countings of cross-ratios mentioned above   -- with and without  Gram determinant
conditions  taken into account -- predict that  no cross-ratios can be written for $n=4,5$,
whereas  at six points they allow for three independent harmonic ratios.
These could be chosen to be
\beq
\label{sixptcr}
u_{36}  \, = \,  {x_{31}^2 x_{46}^2 \over x_{36}^2 x_{4 1}^2} \, := \,  u_1 \ , \qquad
u_{1 4} \, = \,   {x_{15}^2 x_{24}^2 \over x_{14}^2 x_{2 5}^2} \, := \,  u_2
\ , \qquad
u_{25} \, = \,  {x_{26}^2 x_{35}^2 \over x_{25}^2 x_{3 6}^2} \, := \,  u_3
\ .
\eeq
In \cite{seven, dhks6} it was verified for several kinematical configurations that the six-point remainder function
indeed depends on the kinematics only through the three cross-ratios in \eqref{sixptcr},
as predicted by dual conformal invariance. Furthermore, the six-point remainder function is
a symmetric function of the three cross-ratios \eqref{sixptcr} \cite{dhksbum}.
It can easily be shown that any permutations  of the three arguments of the
remainder function corresponds to a cyclic relabeling of the dual momenta plus possibly a reversal of their ordering,
which clearly leaves the Wilson loop unchanged.
This property  was checked numerically in  \cite{seven, dhks6}.

Interestingly, in \cite{dhks6} it was also shown numerically in a  few  examples  that two kinematical configurations which have the same cross-ratios, but differ in that one respects the Gram determinant constraint and one violates it, give rise to the same numerical values for the remainder function.
In this paper we perform explicit calculations of lightlike Wilson loops at six, seven and eight points.
Starting from seven points, we find that there is a different number of cross-ratios depending on whether
one implements the Gram determinant constraint or not.

In practice,  in the following we will discard the Gram constraint altogether,
and work with unconstrained kinematics. This turns out to be a particularly efficient
way to generate kinematical points -- including cases where $n$ is odd.
At seven points, we will  therefore
consider seven (rather than six)
cross-ratios of the form  \eqref{ourcrossratios}.
At eight points, we will consider twelve cross-ratios (rather than nine).
In all cases, we have performed extensive numerical checks proving that
the remainder function only depends on kinematics through the expected
cross-ratios.

\section{Wilson loops and scattering amplitudes}
The Wilson loop we consider in this paper is purely bosonic,
and its expression is given by
\beq
\label{wil}
W[ \cC_n]  \ := \ {\rm Tr} \, \cP \exp \left[ i g\oint_{\cC_n} \! d\tau  \ \dot{x}^{\mu} (\tau )A_\mu (x(\tau ))   \right]
\ .
\eeq
The particular closed contour $\cC_n$ we consider is the lightlike $n$-edged polygonal contour introduced in \cite{am}.
It is obtained by  attaching the momenta of the scattered particles $p_1, \ldots , p_n$ one after the other, following the order
of the colour generators in the colour-ordered scattering amplitude. The resulting contour is closed as
$\sum_{i=1}^n p_i = 0$, and the positions of the vertices are given by the dual momenta coordinates, introduced earlier in
\eqref{dm}.

Calculations of  \eqref{wil} at one loop were performed in \cite{dks} and \cite{bht}, where agreement was found with the
expression of the scalar function in \eqref{Mfunction} appearing in the corresponding one-loop
MHV amplitude.
In the following we will discuss the basic ingredients needed to perform a two-loop perturbative calculation of the
Wilson loop.

\subsection{Perturbation theory setup}
The calculation of the Wilson loop at higher loops is simplified if one makes use of the non-abelian exponentiation theorem
\cite{gatheral,taylor}. This theorem allows one to write the result of the vacuum expectation value of the Wilson loop
as an exponential, and gives a practical rule to calculate the exponent.
We represent the Wilson loop as
\beq
\label{nae}
\lan W[\cC_n ] \ran \  :=  1 \, + \, \sum_{l=1}^{\infty} a^l W^{(l)}_n \ := \   \exp \sum_{l=1}^{\infty} a^l w^{(l)}_n
\ ,
\eeq
and, in this paper,  our main focus is on the evaluation of the two-loop term $w^{(2)}$ in the exponent. In terms of the Wilson loop
coefficients $W^{(l)}_n$ this is
obtained as%
\footnote{Our Wilson loop conventions are summarised and compared to  those of \cite{dhks} in Appendix A.}
\beq
\label{w2our}
w^{(2)}_n \ = \  W^{(2)}_n \, - \, {1\over2} \, (W^{(1)}_n)^2
\ .
\eeq
The non-abelian exponentiation theorem
has been used widely in several Wilson loop calculations, see for example
\cite{kk,km}, and more recently \cite{dhks4,dhks5,dhks6}. To briefly illustrate its application,
we first consider the calculation of a Wilson loop vacuum expectation value  in an abelian theory.
In this case, it is not difficult to see that the perturbative series
reorganises itself into the exponential of the one-loop correction, i.e.~the corresponding abelian result
is given by a formula like \eqref{nae} with $w_{\mathrm{\small QED}}^{(l)} =0$, for any  $l >1$.
In the non-abelian case, parts of the result of the diagrams contribute to the exponentiation of the one-loop result,
but there are additional contributions  which correct order by order in perturbation theory
the one-loop term in the exponent.
In brief, the rule for calculating the complete exponent \cite{gatheral, taylor} is to restrict
to those parts of the diagrams which give a ``maximal non-abelian colour factor".
At two loops, this turns out to be equal  to $C_F C_A$ \cite{km}, where
$ C_F :=  C_2 (\mathbf{r} ) $ is the Casimir in the representation $\mathbf{r}$ of the Wilson loop,  and
$ C_A  :=  C_2 (\mathbf{G} ) $ is the adjoint Casimir.%
\footnote{We notice that, in order to be properly normalised, the Wilson loop in \eqref{wil} should be divided
by the dimension of the representation $d_F := d(\mathbf{r})$.}
For $SU(N)$, one has $C_F= (N^2-1)/(2N)$ for the fundamental representation, and $C_A=N$.

As a simple example, consider the two-loop contribution to a cusp diagram arising from diagrams containing
only propagators.  The contribution from a ladder diagram produces the colour factor
$\mathrm{Tr} (T^a T^a T^b T^b) = d_F C_F^2$, whereas the
cross propagator diagram, represented on the left hand side of Figure \ref{fig:cusps},  contains the colour factor
$\mathrm{Tr} (T^a T^b T^a T^b) =d_F C_F ( C_F - 1/2 \,C_A) $. According to the
non-abelian exponentiation theorem, we only have to consider the term
$-(1/2) \, C_F C_A$ from the cross propagator diagram, and discard the remaining diagram altogether (which has already
been taken contributed to the exponentiation of the one-loop correction).

As a final remark, we would like to observe that the diagrams needed to calculate these maximally non-abelian
corrections  are simpler (and fewer) than those needed for the full Wilson loop, however the technical difficulties
in obtaining the final integrals in analytic form are typically  comparable.

We now move on to describing the basic ingredients of any Wilson loop  perturbative calculation.
The first one is the gluon propagator which,
in the Feynman  gauge, is given by
$\D_{\mu \nu} (x) := \eta_{\m \n} \D(x)$, where
\beqa
\label{delta}
\D  (x) & := &
- {\pi^{2 - {D\over 2}} \over 4 \pi^2}
\Gamma \Big({D\over 2} - 1 \Big)
 {1\over (-x^2+ i \varepsilon)^{{D\over 2} - 1}}
\\ \nonumber
&=&
- {\pi^{\epsuv} \over 4 \pi^2} {
\Gamma (1-\epsuv)  \over
(-x^2+ i \varepsilon)^{1-\epsuv }}
\ ,
\eeqa
where $D=4-2 \epsuv$. The Wilson loop is gauge invariant,  therefore we can pick any gauge we like
to compute its expectation value.%
\footnote{The advantage of considering different gauges, still belonging to the class of Feynman-'t Hooft gauges,
has been discussed recently in \cite{bhnt}. A different possibility would be to pick the  lightcone gauge.
This gauge has been used for  Wilson loop calculations  in \cite{Bassetto:1993xd}.}

At two loops, we will also have Feynman diagrams where the gluon three-point vertex contributes.
The basic structure to know is therefore the Wick contraction of three gauge fields  with a
three-point vertex,
\beqa
\nonumber
&&
\mathrm{Wick} \Big[
A^{\m_1, a_1}  (x_1)  A^{\m_2, a_2} (x_2)  A^{\m_3, a_3}  (x_3)
\int\!\! d^D z \  \mathrm{Tr}  \Big( \partial_\m A_\n [ A^\m, A^\n] \Big)(z)
\Big]  \ = \  -i C_F f^{a_1 a_2 a_3} \times
\\
&&
 \ \Big[ \eta^{\mu_1 \mu_2} (\del^{\mu_3}_1 - \del^{\mu_3}_2 )  +
\eta^{\mu_2 \mu_3} (\del^{\mu_1}_1 - \del^{\mu_1}_2 ) + \eta^{\mu_3 \mu_1} (\del^{\mu_2}_1 - \del^{\mu_2}_2 )
\Big] G(x_1, x_2, x_3)
\ ,
\eeqa
where
\beqa
\label{G}
&&G(x_1, x_2, x_3) \ := \
\int\!\! d^Dz \ \Delta (x_1 -z) \Delta (x_2 -z) \Delta (x_3 -z)
\\
&&\,  = \,
{(i)^{1-2D}\over 64 \pi^D }
\Gamma(D-3)
\int\!\!\prod_{i=1}^{3} d \a_i \
\delta(1 - \sum_{i=1}^3 \a_1 )
{
(\a_1 \a_2 \a_3 )^{{D\over 2} - 2} \over
(\a_1 \a_2 x_{12}^2 + \a_1 \a_3 x_{13}^2 + \a_2 \a_3 x_{23}^2 )^{D-3}
}
\ ,
\nonumber
\eeqa
where we have used  \eqref{delta} and $x_{ij}^2 = (x_i - x_j)^2$.
The evaluation of the right hand side of \eqref{G} in various cases, specifically when
$x_2^2 = x_3^2 = x_{23}^2 = 0$,  has been carried out in \cite{km}.

Finally, we notice that the colour factor associated with gluon three-point vertex diagrams,
obtained after contracting with a trace of three colour generators,
is $\mathrm{Tr} ( T^{a_1} T^{a_2}T^{a_3}) f^{a_1 a_2 a_3} =  (1/2) d_F C_F C_A$.

\subsection{The Wilson loop remainder function}
\label{WLremainderfunction}

We define the $n$-sided Wilson loop remainder function $\cR_n^\mathrm{WL} $
in complete analogy  with the amplitude remainder function  introduced in
\eqref{fm2}, as
\begin{align}
\label{wbds}
  w_n^{(2)}(\e)  &= f_\mathrm{WL}^{(2)}(\epsilon) \, w_n^{(1)}(2 \epsilon) \, + \, C_\mathrm{WL}^{(2)} \, +\, \cR_n^\mathrm{WL}
  \  ,
\end{align}
where $\eps=-\epsuv$.
We have added a subscript WL to distinguish quantities relevant for the Wilson loop from the corresponding amplitude expressions.
In particular,
\beq \label{flepsW}
f_\mathrm{WL}^{(2)}(\epsilon ) :=\, f_0^{(2)} + f_{1,\mathrm{WL}}^{(2)} \epsilon + f_{2,\mathrm{WL}}^{(2)} \epsilon^2
\ ,
\eeq
where $f_0^{(2)}$  is the same as in \eqref{fleps}, whilst
$f_{1,\mathrm{WL}}^{(2)}= G_{\rm eik}^{(2)} = 7 \zeta_3 $  \cite{kk},
and the third term $f_{2,\mathrm{WL}}^{(2)}$ is a so far undetermined constant.
Similarly, the constant $C^{(2)}_\mathrm{WL}$ in  \eqref{fm2} has also not been fixed yet. We will shortly determine
these two constants.

A few comments are in order.

Firstly, we notice the already mentioned discrepancy between   the coefficient $ G_{\rm eik}^{(2)}$ of the subleading $1/\e$ pole in the Wilson loop and the corresponding coefficient $G^{(2)}$ appearing in \eqref{defGcoll} on the amplitude side.
This discrepancy has been examined and explained in \cite{dms}.
We note that this discrepancy
cannot be reabsorbed into a (kinematic-independent) redefinition of the Wilson loop
renormalisation scale $\mu_\mathrm{WL}$ alone.%
\footnote{See also the discussion in \cite{dhks6} (version 3). }

Secondly, we would like to point out that if the correct determination of the  constants  $ f_{2,\mathrm{WL}}^{(2)}$ and
$C^{(2)}_\mathrm{WL}$ is implemented also for Wilson loops,
we expect the Wilson loop remainder function to have precisely the same collinear limit as
its amplitude counterpart,   namely
\begin{equation}
\label{abb}
  \cR_n^\mathrm{WL} \rightarrow \cR_{n-1}^\mathrm{WL} \ ,
\end{equation}
with no extra constant term on the right hand side of \eqref{abb}.

In order to determine $C^{(2)}_\mathrm{WL}$ and $f_{2, \mathrm{WL}}^{(2)}$, and later be able to
check \eqref{abb},  we proceed as follows.
Firstly, we recall that conformal invariance guarantees that the four- and five- point Wilson loops
satisfy an ABDK/BDS-like ansatz \cite{dhks5}. This implies
that the remainders $\cR_4^\mathrm{WL}$ and $\cR_5^\mathrm{WL}$ cannot depend on kinematics and must be constant.
On the amplitude side, these remainder functions are known to vanish. Thus we also choose
\beq
\label{R45}
\cR_4^\mathrm{WL}\, = \, \cR_5^\mathrm{WL} \, = \, 0
\ ,
\eeq
in \eqref{wbds} and  then determine
 $C^{(2)}_\mathrm{WL}$ and $f_{2, \mathrm{WL}}^{(2)}$ from solving \eqref{wbds} for $n=4,5$.

Notice that in writing the Wilson loop ABDK/BDS ansatz, it is crucial to use the
one-loop Wilson loop, and not the one-loop
amplitude. The two are equal to all orders in $\e$  up to their
normalisation~\cite{bht}. More concretely, for the amplitude we have
$\cM_n^{(1)}= 2 \hat c_\Gamma
\cM_{n,\mathrm{BDDK}}^{(1)}$ where $\cM_{n,\mathrm{BDDK}}^{(1)}$ is
the one-loop amplitude in the normalisations of\cite{bddk}, and
\begin{align}\label{cg}
\hat{c}_\G \ := \ {e^{\e \gamma} \over 2} {\G (1 + \e) \G^2 ( 1 -  \e) \over
\G(1 - 2 \e)}
\ .
\end{align}
This leads to $ \cM_n^{(1)}=
e^{\epsilon\gamma} \Gamma(1+\epsilon)
(\Gamma(1-\epsilon)^2/\Gamma(1-2\epsilon)) \cM_{n,\mathrm{BDDK}}^{(1)}
$. On the other hand, for the
Wilson loop, we have%
\footnote{In the following formulae we employ the  redefinition of the renormalisation scale in \eqref{muWL}.}
$w_n^{(1)} = e^{\epsilon \gamma} \Gamma(1+\epsilon)
 \cM_{n,\mathrm{BDDK}}^{(1)}$.
 This leads to the following correspondence between the Wilson loop
 and the amplitude at one loop,
\begin{equation}\label{aw}
  w_n^{(1)}\ = \  {\G(1 - 2 \e)  \over  \G^2 ( 1 -  \e)} \cM_n^{(1)} \ = \ (1+\zeta_2
  \epsilon^2) \cM_n^{(1)}\, + \, \cO (\e) \ = \  \cM_n^{(1)}  \, - \, n\,  { \pi^2 \over 12}\, + \, \cO (\e)
  \ .
\end{equation}
At one loop,  the four- \cite{dks} and five-edged Wilson loops
\cite{bht} are thus given by
\begin{align}
  w^{(1)}_4&= -
  {1\over \e^2}
  \left[
  \left(
 - {s \over \mu_\mathrm{}^2}
  \right)^{-\e}  +
 \left( - {t \over \mu_\mathrm{}^2}\right)^{-\e}
  \right]
  \ + \ {1\over 2} \log^2 \left( s\over
t\right) + {\pi^2 \over 3} \ ,\\
  w^{(1)}_5&=  {1\over 2} \sum_{i=1}^5  \left[- {1\over \e^2}
  \left( - {t_i^{[2]} \over \mu_\mathrm{}^2 }\right)^{-\e}
   -{1\over 2} \ln  \left( { -t^{[2]}_{i}\over -t^{[3]}_{i} }\right)
  \ln \left({ -t^{[2]}_{i+1}\over -t^{[2]}_{i+2} }\right)   +
  {\pi^2\over 12}\right]\ ,
\end{align}
and at two loops~\cite{dhks4,dhks5}
\begin{align}
\label{eq:1}
  w^{(2)}_4 &=   {2 } \left[
  \left(
 - {s \over \mu_\mathrm{}^2}
  \right)^{-2\e}  +
 \left( - {t \over \mu_\mathrm{}^2}\right)^{-2\e}
  \right] \left( {\pi^2\over 48 \epsilon^2} -{7 \zeta_3
    \over 8\epsilon}  \right)  - {\pi^2 \over 12}
\log^2 \left( s\over t\right)   -{\pi^4\over 24} \ ,\\
  w^{(2)}_5&=  \sum_{i=1}^5   \left( - {t_i^{[2]} \over \mu_\mathrm{}^2 }\right)^{-2\e}
  \left(  {\pi^2\over 48 \epsilon^2} -{7 \zeta_3
    \over 8\epsilon}   \right)                  +{\pi^2 \over 24}
\sum_{i=1}^5 \ln
\left( { -t^{[2]}_{i}\over -t^{[3]}_{i} }\right)
  \ln \left({ -t^{[2]}_{i+1}\over -t^{[2]}_{i+2} }\right)
  - {\pi^4\over 72} \ .
  \label{eq:2}
\end{align}
We note  that in  \eqref{eq:1} we have used the results of our two-loop calculation of the four-point Wilson loop
to correct the constant term in  the corresponding  result  of \cite{dhks4}.%
\footnote{This discrepancy has also been
noted independently by Marcus Spradlin, whom we thank for discussions on this point.}

We can now uniquely rewrite  \eqref{eq:1}  and \eqref{eq:2}
in an ABDK/BDS form as
\begin{align}
  w_4^{(2)} (\e) &= f_\mathrm{WL}^{(2)}(\epsilon) \, w_4^{(1)}(2 \epsilon) \, + \, C_\mathrm{WL}^{(2)}\ ,
  \\
  w_5^{(2)}(\e)  &= f_\mathrm{WL}^{(2)}(\epsilon) \, w_5^{(1)}(2 \epsilon) \, + \, C_\mathrm{WL}^{(2)}
\ ,
\end{align}
where
\beq
f_\mathrm{WL}^{(2)}(\epsilon)\ =\  -\zeta_2 \, + \,   7 \zeta_3 \, \e  \ - \    5\zeta_4\, \e^2
\ ,
\eeq
and
\beq
C_\mathrm{WL}^{(2)} \ =\  -{1\over 2} \zeta_2^2
\ .
\eeq
The $\cO (1)$ and $\cO (\e )$ coefficients of $f_\mathrm{WL}^{(2)}(\epsilon)$ had already been determined in
\cite{dhks4}.
Interestingly,  the constant $C_\mathrm{WL}^{(2)}$ turns out to be
the  same as the constant $C^{(2)}$ in \eqref{c2} for the amplitude.

Finally, let us now compare the definition of the remainder function $\cR_6^\mathrm{WL}$ given in \eqref{wbds}
with that  of DHKS~\cite{dhksbum,dhks6} (see also Appendix A).
First, we write the two-loop term $w^{(2)}$ in the form
\beq
\label{no2}
w^{(2)}_n \ = \
\sum_{i=1}^n { \left( -{ x_{i i+2}^2\over \m^2}\right)^{-2\eps} } \Big(
{ w^{(2)}_{-2} \over \eps^2} +
{w^{(2)}_{-1} \over \eps}  \Big)  \ + \ F^{(2)}_{n} \, + \, \cO(\e )
\ ,
\eeq
where $F^{(2)}_{n}$ is finite as $\e \to 0$.

{}From \eqref{eq:1} and \eqref{eq:2}, one has
\beq
w^{(2)}_{-2} \, = \, {\pi^2 \over 48} \ \  ,
\qquad
w^{(2)}_{-1} \, = \, - {7\over 8 }\zeta_3
\ ,
\eeq
and comparing \eqref{no2} with \eqref{wbds}, using \eqref{aw} and \eqref{m1}, we obtain
\begin{align}
  F_n^{(2)}\ = \  {1\over 4} \gamma_K^{(2)}\, \, F^{(1)}_n (0)  \, +\,
  C_\mathrm{WL}^{(2)} +\cR_n^\mathrm{WL} \, +
  \, n\, {\pi^4\over 48} \ .
\end{align}
The DHKS finite remainder function is then defined as \cite{dhks6,seven}
\beq
\label{remainder}
\cR_n^\mathrm{DHKS}  \ :=  \ F^{(2)}_{n} \ - \ F^{\mathrm{BDS}\, (2)}_n \ ,
\eeq
where $F^{\mathrm{BDS}\, (2)}_n$ is the two-loop contribution to $F^\mathrm{BDS}_n$  in \eqref{fbds},
\beq
\label{fbds2}
F^{\mathrm{BDS}\, (2)}_n  = \, {1\over 4} \gamma_K^{(2)}\, \, F^{(1)}_n (0)  + C^{(2)}
\ .
\eeq
Thus we find that
our finite remainder defined in \eqref{wbds} and the DHKS definition \eqref{remainder}
are related by a constant shift,
\beq
\label{sshift}
\cR_n^\mathrm{WL}\, = \, \cR_n^{\mathrm{DHKS}} \, -\,  n\, { \pi^4 \over 48}\ .
\eeq
We have checked that
our Wilson loop remainder function $\cR_n^\mathrm{WL}$ satisfies
\eqref{abb} under a collinear limit.
For $n=6$ this amounts to
 $\cR_6^\mathrm{WL}\rightarrow \cR_5^\mathrm{WL} =0$.
This would imply that
\beq
\cR_6^{\mathrm{DHKS}} \rightarrow \pi^4 / 8 \, =\, 12.1761...
\ ,
\eeq
which is completely consistent with the results
of~\cite{dhks6}, where this was computed  numerically as $c_\mathrm{W}  := 12.1756$ with
accuracy of the order $10^{-3}$.
In the following sections we will show that \eqref{abb} also holds for $n=7, 8$.

We can now express
the statement of the duality between Wilson loops and amplitudes as an
equality of the corresponding remainder functions defined in \eqref{wbds} and \eqref{fm2},
\begin{align}
\label{dualitywa}
 \cR_n \ = \   \cR_n^\mathrm{WL}\ .
\end{align}
Notice that no additional constant term is allowed on the right hand side of
\eqref{dualitywa}.


\section{Summary of the diagrams entering the Wilson loop at two loops }
\label{sec:wilsondiags}
\begin{figure}[h]
\begin{center}
\scalebox{0.55}
{
\fcolorbox{white}{white}{
  \begin{picture}(448,309) (136,-93)
    \SetWidth{0.5}
    \SetColor{Black}
    \ArrowLine(420,-21)(289,-21)
    \ArrowLine(225,-5)(225,131)
    \ArrowLine(495,129)(495,-5)
    \ArrowLine(285,144)(420,144)
    \Gluon(353,144)(354,-20){7.5}{10.29}
    \Gluon(224,62)(340,61){7.5}{6.86}
    \Gluon(368,61)(495,61){7.5}{7.64}
    \Text(136,58)[lb]{\Large{\Black{$p_i$}}}
    \Text(554,58)[lb]{\Large{\Black{$p_j$}}}
    \Text(345,200)[lb]{\Large{\Black{$p_k$}}}
    \Text(345,-93)[lb]{\Large{\Black{$p_l$}}}
  \end{picture}
}
}
\end{center}
 \caption{\it The factorised cross diagram.}
\label{factorisedcrossdiagram}
\end{figure}
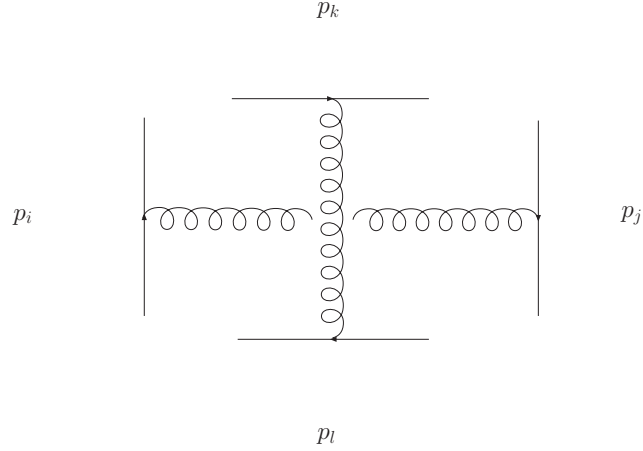

In this section we summarise the expressions for all the diagrams
entering a generic
two-loop Wilson loop calculation.
There are five main ingredients to the two-loop Wilson loop
calculation at any number
of edges, $n$. We call them the ``hard diagram'', the ``curtain diagram'',
the ``cross diagram'',  the ``Y diagram'', and the  ``factorised cross
diagram", see Figures \ref{harddiagram}--\ref{factorisedcrossdiagram}.

In the following we summarise the final expressions for the integrals
corresponding to these diagrams; derivations are
outlined in the Appendices.
The entire two-loop contribution to the logarithm of the Wilson loop is assembled in terms of the individual building blocks
in Section \ref{sec:complete-Wilson}.

In all expressions of the diagrams we will write in the next sections,
a factor of
\beq
\label{multby}
\cC:=2 a^2 \mu^{4\epsilon} \Big[{\Gamma ( 1 + \e) e^{  \gamma \e}}\Big]^2
\ = \ 2 a^2 \mu^{4\epsilon} \Big
( 1 + {\pi^2 \over 6} \e^2\Big) \, + \, \cO (\e^3 )
\ ,
\eeq
will be pulled out,
where we have  defined the coupling
   \beq
a\  := \  {g^2 N \over  8 \pi^2}
\ ,
\eeq
and the scale, $\mu^2$, is given in terms of the Wilson loop scale as
\beq
\label{muWL}
\mu^2_{\rm WL} \ :=  \  \pi e^{\gamma} \mu^2
\ .
\eeq
This factor will be reintroduced when the diagrams are
reassembled into the complete Wilson loop~(\ref{completewl})
in order to match the conventions of \cite{dhks6}  and facilitate
comparisons with their results.

\subsection{The hard diagram}
The hard diagram is depicted in Figure \ref{harddiagram}, and is given by the integral:
 \begin{align}
 \label{eq:hardexpression}
&  f_H(p_1,p_2,p_3;Q_1,Q_2,Q_3)\nonumber \\
&:= {1 \over 8} \, {\Gamma (2-2\epsuv) \over \Gamma(1-\epsuv)^2} \int_{0}^{1}\!
\Big(  \prod_{i=1}^{3} d\t_i \Big) \int_{0}^{1}\!\Big(\prod_{i=1}^{3} \, d\alpha_i \Big)\delta ( 1 - \sum_{i=1}^3
\alpha_i )  \ (\alpha_1 \alpha_2 \alpha_3)^{-\epsuv}  { \mathcal{N} \over \mathcal{D}^{2-2\epsuv}}\ ,
\end{align}
where the functions $\cD$ and $\cN$ are given in \eqref{D}, \eqref{N}.
The momenta $p_i$ are massless $p_i^2=0$ and the $Q_i$
can be
massive.
They are further constrained by momentum conservation
\begin{align}
\label{asinthe}
  p_1+p_2+p_3+Q_1+Q_2+Q_3=0\ .
\end{align}
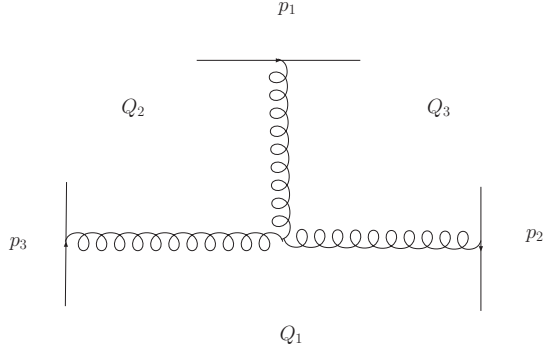
\begin{figure}[h]
\begin{center}
\scalebox{0.45}
{
\fcolorbox{white}{white}{
  \begin{picture}(464,293) (143,-64)
    \SetWidth{0.5}
    \SetColor{Black}
    \ArrowLine(299,174)(436,174)
    \ArrowLine(189,-32)(190,72)
    \ArrowLine(537,68)(537,-36)
    \Gluon(367,175)(370,24){7.5}{9.36}
    \Gluon(189,22)(371,22){7.5}{11.57}
    \Gluon(537,23)(372,25){7.5}{10.36}
    \Text(143,15)[lb]{\Large{\Black{$p_3$}}}
    \Text(369,213)[lb]{\Large{\Black{$p_1$}}}
    \Text(577,22)[lb]{\Large{\Black{$p_2$}}}
    \Text(370,-64)[lb]{\Large{\Black{$Q_1$}}}
    \Text(237,127)[lb]{\Large{\Black{$Q_2$}}}
    \Text(494,127)[lb]{\Large{\Black{$Q_3$}}}
  \end{picture}
}}
\end{center}
\caption{\it The hard diagram.}
\label{harddiagram}
\end{figure}

\subsection{The curtain diagram}

\begin{figure}[h]
\begin{center}
\scalebox{0.47}{
\fcolorbox{white}{white}{
  \begin{picture}(464,293) (143,-64)
    \SetWidth{0.5}
    \SetColor{Black}
    \ArrowLine(299,174)(436,174)
    \ArrowLine(189,-32)(190,72)
    \ArrowLine(537,68)(537,-36)
    \Text(143,15)[lb]{\Large{\Black{$p_3$}}}
    \Text(369,213)[lb]{\Large{\Black{$p_1$}}}
    \Text(577,22)[lb]{\Large{\Black{$p_2$}}}
    \Text(370,-64)[lb]{\Large{\Black{$Q_1$}}}
    \Text(237,113)[lb]{\Large{\Black{$Q_2$}}}
    \Text(494,127)[lb]{\Large{\Black{$Q_3$}}}
    \GlueArc(469.98,122.97)(151.15,161.07,295.9){-7.5}{23.98}
    \GlueArc(249.42,147.62)(167.51,-110.78,8.71){-7.5}{23.53}
  \end{picture}
}
}
\end{center}
\caption{\it The curtain diagram.}
\label{curtaindiagram}
\end{figure}
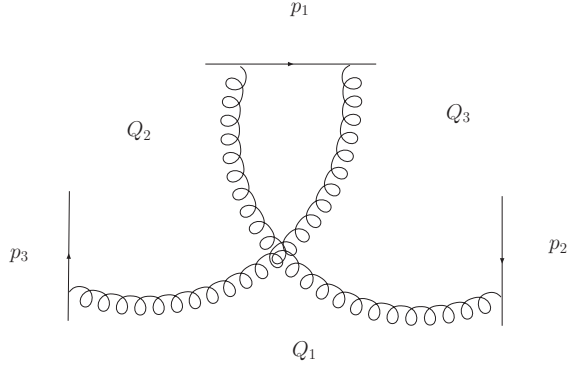

The curtain diagram is represented in Figure \ref{curtaindiagram}, and its expression is
given by the integral
\beqa
&&
f_C(p_1,p_2,p_3;Q_1,Q_2,Q_3)
\nonumber \\
&&:=   \ -
{1\over 2} \,
\int_0^1\! \Big(  \prod_{i=1}^{3} d\t_i
\Big)\,\int_{1-\tau_1}^{1}d\s_1 {(p_1p_2) \over \big[ -2(p_1Q_3)
  \s_1-2(p_1p_2)\s_1 \t_2-2(p_2Q_3)\t_2-Q_3^2 \big]^{1-\epsuv}}
  \nonumber \\
&&
\times {(p_1 p_3) \over
 \big[  -2(p_1Q_2) \t_1-2(p_1p_3)\t_1\t_3-2(p_3Q_2)\t_3-Q_2^2
\big]^{1-\epsuv}}
\ .
\label{curtain}
\eeqa
The $p_i$ and $Q_i$ are constrained as in \eqref{asinthe}.

\subsection{The cross diagram}

The cross diagram is represented in Figure \ref{crossdiagram}, and  is given by
\begin{align}
  &f_X(p_1,p_2;Q_1,Q_2)\nonumber \\
&:=-{1\over 2} \, \int_0^1 d \sigma_1 d\tau_2
 \int_0^{\sigma_1}\!\!d\tau_1 \int_0^{\tau_2}\!\!d\sigma_2 &{(p_1p_2) \over
   \big(-2(p_1p_2)\sigma_1\sigma_2-2p_1Q_2\sigma_1-2p_2Q_2\sigma_2-Q_2^2
   \big)^{1-\epsuv}}\nonumber
\\
 &&{(p_1p_2) \over
   \big(-2(p_1p_2)\tau_1\tau_2-2p_1Q_2\tau_1-2p_2Q_2\tau_2-Q_2^2\big)^{1-\epsuv}}
\end{align}
Again the $p_i$ are massless and the $Q_i$ massive and
momentum conservation is imposed,
\begin{align}
  p_1+p_2+Q_1+Q_2=0\ .
\end{align}
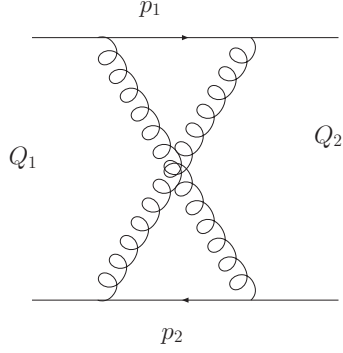
\begin{figure}[h]
\begin{center}
\scalebox{0.55}{
\fcolorbox{white}{white}{
  \begin{picture}(240,241) (210,-59)
    \SetWidth{0.5}
    \SetColor{Black}
    \ArrowLine(225,151)(435,151)
    \ArrowLine(435,-29)(225,-29)
    \Gluon(270,151)(375,-29){7.5}{13.46}
    \Gluon(375,151)(270,-29){7.5}{13.46}
    \Text(300,166)[lb]{\Large{\Black{$p_1$}}}
    \Text(420,76)[lb]{\Large{\Black{$Q_2$}}}
    \Text(210,61)[lb]{\Large{\Black{$Q_1$}}}
    \Text(315,-59)[lb]{\Large{\Black{$p_2$}}}
  \end{picture}
}
}
\end{center}
\caption{\it The cross diagram.}
\label{crossdiagram}
\end{figure}

\subsection{The Y diagram + self-energy diagram}

\begin{figure}[t]

\begin{center}
\scalebox{.55}{
\fcolorbox{white}{white}{
 \begin{picture}(584,250) (40,-61)
   \SetWidth{0.5}
   \SetColor{Black}
   \Gluon(107,160)(167,69){7.5}{6.06}
   \Gluon(166,69)(226,160){7.5}{6.06}
   \Gluon(167,-21)(167,69){7.5}{5}
   \Text(165,173)[lb]{\Large{\Black{$p_1$}}}
   \Text(40,80)[lb]{\Large{\Black{$Q_1$}}}
   \Text(279,80)[lb]{\Large{\Black{$Q_2$}}}
   \ArrowLine(273,-20)(63,-20)
   \ArrowLine(61,160)(270,160)
   \ArrowLine(403,160)(613,160)
   \ArrowLine(615,-20)(405,-20)
   \Gluon(505,-20)(505,57){7.5}{3.49}
   \Gluon(504,85)(504,160){7.5}{2.57}
   \SetColor{Blue}
   \Vertex(503,70){20.81}
   \Text(510,173)[lb]{\Large{\Black{$p_1$}}}
   \Text(165,-61)[lb]{\Large{\Black{$p_2$}}}
   \Text(510,-61)[lb]{\Large{\Black{$p_2$}}}
   \Text(375,80)[lb]{\Large{\Black{$Q_1$}}}
   \Text(594,80)[lb]{\Large{\Black{$Q_2$}}}
 \end{picture}
}
}
\end{center}
  \caption{\it The Y diagram together with the self-energy diagram.}
\label{Ydiagram}
\end{figure}
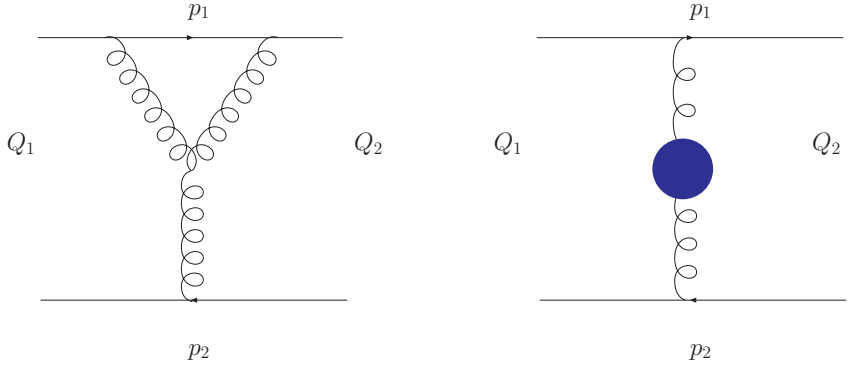

The Y diagram, to which we also add (half of) the self-energy diagram,\footnote{The other half
of the self-energy accompanies the ``upside-down" Y diagram. }
represented in Figure \ref{Ydiagram}, is given by the following
integral,
\beqa
&& f_Y(p_1,p_2;Q_1,Q_2)
\ :=  \
{p_1 \cdot p_2\over 8 } {1\over \epsuv}
    {\Gamma(1-2\epsuv) \over \Gamma^2(1-\epsuv)}
    \\ \nonumber
    && \qquad
    \int_0^1\!\!d\sigma
    \int_0^1\!\!d\tau_1  d\tau_2
    \Big[-{\sigma^{-\epsuv}(1-\sigma)^{-\epsuv} \over (-
      Q_1^2-2 (Q_1p_2)
    \tau_2 -
  2(Q_1p_1) \sigma \tau_1-2(p_1p_2) \sigma \tau_1 \tau_2)^{1-2\epsuv}} \nonumber \\
 && \qquad \qquad\qquad
  -{\sigma^{-\epsuv}(1-\sigma)^{-\epsuv} \over (-Q_2^2-2(Q_2p_2)\tau_2-
   2(Q_2p_1)\sigma \tau_1 -2(p_1p_2)
  \sigma \tau_1\tau_2 )^{1-2\epsuv}}\Big] \ .
  \nonumber
\eeqa

\subsection{The factorised cross diagram}

This is given as $-1/2$ times the product of two one-loop diagrams
\begin{align}
-1/2  f_{P}(p_i,p_j;Q_{ji},Q_{ij})
f_P(p_k,p_l;Q_{lk},Q_{kl}) \ .
\end{align}
The one-loop terms come from a diagram involving just a
single propagator, and
are simply the finite part of the two-mass easy box
function~\cite{bht}
\begin{align}
f_P(p,q;P,Q)&=
{1\over 2}\bigg[ \frac{1}{2} \log
    ^2\left(\frac{s}{t}\right)+\text{Li}_2\left(1-\frac{P^2}{s}\right)+\text{Li}_2\left(1-\frac{Q^2}{s}\right)+\text{Li}_2\left(1-\frac{P^2}{
    t}\right)\nonumber \\
& \qquad \qquad +\text{Li}_2\left(1-\frac{Q^2}{t}\right)-\text{Li}_2\left(1-\frac{P^2
    Q^2}{s t} \right)
 \bigg]
 \ ,
\end{align}
where
$s=(  P+p)^2$ and $t=(P+q)^2$.

\subsection{The complete Wilson loop  at  $n$ points}\label{sec:complete-Wilson}

The logarithm  of the complete $n$-sided Wilson loop is given by the sum over all diagrams,
\begin{align}  w_n^{(2)}= \cC \
\bigg\{& \sum_{1\leq i<j<k\leq n}
  \Big[f_H(p_i,p_j,p_k;Q_{jk},Q_{ki},Q_{ij})+ f_C(p_i,p_j,p_k;Q_{jk},Q_{ki},Q_{ij})
  \nonumber \\
&\qquad \qquad \qquad +
  f_C(p_j,p_k,p_i;Q_{ki},Q_{ij},Q_{jk})+
  f_C(p_k,p_i,p_j;Q_{ij},Q_{jk},Q_{ki}) \Big]\nonumber \\[10pt]
&+  \sum_{1\leq i<j \leq n} \Big[ f_X(p_i,p_j;Q_{ji},Q_{ij}) +
f_Y(p_i,p_j;Q_{ji},Q_{ij}) +f_Y(p_j,p_i;Q_{ij},Q_{ji})\Big]\nonumber\\[10pt]
&+  \sum_{1\leq i<k<j<l \leq n}  (-1/2) f_{P}(p_i,p_j;Q_{ji},Q_{ij})
f_P(p_k,p_l;Q_{lk},Q_{kl})\bigg\}
\ , \label{completewl}
\end{align}
where the first sum is over all sets of three non-equal legs $i,j,k$,
the second sum is over all sets of two non-equal legs $i,j$, and the
third sum over all sets of four non-equal legs.
Here
we have defined
\begin{align}
  Q_{ij}=p_{i+1}+p_{i+2}+\dots p_{j-1}\ ,
\end{align}
and
$\cC$
is the factor we pulled out (see~\eqref{multby}).

Note that the singular properties of these integrals depends on
whether $Q_i=0$ or not (i.e.~whether legs are adjacent or not). For
example $f_H$ has a $1/\epsuv^2$ singularity if $Q_1=Q_2=0,\ Q_3
\neq0$, a $1/\epsuv$ singularity if $Q_1=0,\ Q_2,Q_3 \neq 0$, and is
finite as $\epsuv \rightarrow 0$ if $Q_1,Q_2,Q_3 \neq0$.

In the four-point case, for example, \eqref{completewl} leads
to
\begin{align}
w^{(2)}_4= \cC \bigg\{ &f_H(p_1,p_2,p_3;0,p_4,0) \nonumber\\
&+f_C(p_1,p_2,p_4;p_3,0,0)+f_C(p_1,p_2,p_3;0,p_4,0)+f_C(p_1,p_3,p_4;0,0,p_2)
\nonumber\\
&+{1\over 2} f_X(p_1,p_3;p_4,p_2)+f_X(p_1,p_2;p_1+p_2,0)\nonumber\\
&+f_Y(p_1,p_3;p_4,p_2)+f_Y(p_1,p_2;p_1+p_2,0)+f_Y(p_2,p_1;0,p_1+p_2)\nonumber\\
&+(-1/8) f_P(p_1,p_4;p_2) f_P(p_2,p_4;p_1,p_3)\nonumber\\
&+\ \mathrm{cyclic\  permutations\  of\  }(p_1,p_2,p_3,p_4)  \bigg
\}
\end{align}
(the factor of 1/2 in front of $f_X$ and the extra factor of $1/4$ in
from of the factorised cross is to account for the double counting
of diagrams when summing over cyclic permutations).
Of course everything should only depend on $s=(p_1+p_2)^2$ and $t=(p_1+p_4)^2$.

\subsection{Cusp diagrams}

\begin{figure}[h]
\begin{center}
\scalebox{0.60}{
\fcolorbox{white}{white}{
  \begin{picture}(614,154) (40,-135)
    \SetWidth{0.5}
    \SetColor{Black}
    \ArrowLine(40.01,16.37)(107.31,-135.5)
    \ArrowLine(286.46,19.1)(353.76,-132.77)
    \ArrowLine(491.99,17.28)(559.29,-134.59)
    \ArrowLine(206.44,-50.93)(40.92,16.37)
    \ArrowLine(450.16,-53.66)(286.46,19.1)
    \ArrowLine(654.77,-55.47)(492.9,17.28)
    \Gluon(347.39,-119.13)(350.12,-55.47){6.82}{3.43}
    \Gluon(99.13,-117.31)(101.85,-8.18){6.82}{7.15}
    \Gluon(67.3,-44.56)(186.43,-42.74){6.82}{7.93}
    \Gluon(350.12,-55.47)(411.96,-35.47){6.82}{3.43}
    \Gluon(306.47,-25.46)(350.12,-55.47){6.82}{3.43}
    \Gluon(600.21,-72.75)(631.13,-44.56){6.82}{2.57}
    \SetColor{Blue}
    \Vertex(591.12,-82.76){13.85}
    \SetColor{Black}
    \Gluon(551.1,-114.59)(582.93,-92.76){6.82}{2.57}
  \end{picture}
}
}
\end{center}

\caption{{\it Maximally non-abelian Feynman diagrams contributing to the two-loop cusp corrections. The second diagram
appears with its mirror image where two of the gluon legs of the three-point vertex are attached to the other edge; these two diagrams are equal. The blue bubble in the third diagram represents the gluon self-energy correction calculated in dimensional reduction.  }}
\label{fig:cusps}
\end{figure}
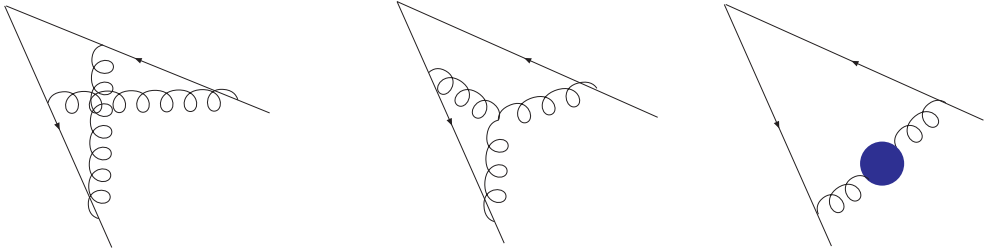

The formula for the exponent of the Wilson loop at two
loops~(\ref{completewl}) includes all contributing diagrams. A subset
of these diagrams involve only two consecutive edges and are known as
cusp diagrams. These are given by
\begin{align}
\cC\   \sum_{i=1}^n \Big(f_X(p_i,p_{i+1};Q_{i+1 i},0) +
f_Y(p_i,p_{i+1};Q_{i+1 i},0) +f_Y(p_{i+1},p_i;0,Q_{i+1 i})\Big)
\, ,
\end{align}
and are shown in Figure~\ref{fig:cusps}.

The final result for the two-loop
correction to the cusps is given by
\beq
\label{cusps1}
  C_FC_A \, \left({g^2 \over 4 \pi^2} \right)^2  \Big[
\Gamma ( 1 + \e) \pi^{-\e} \Big]^2 \sum_{i=1}^n  \left(-{x_{i i+2}^2\over  \mu_{\rm WL}^2}\right)^{- 2 \e}
  \, f_{\rm cusp} (\e )
  \ ,
\eeq
where
\beq
f_\mathrm{cusp} (\e ) \ = \
 \,
{1 \over 2}  {1\over 8 \e^4}\Big[  {\Gamma (1 + 2 \e )  \Gamma (1 - \e ) \over \Gamma (1 + \e ) } - 1\Big]
\ = \  {\pi^2\over 48 \e^2 } \,  -\,  {\zeta_3\over 8 \e} \,  + \, {\pi^4\over 160}\ + \   \cO (\e)
\ .
\eeq
This way, one can rewrite \eqref{cusps1} using the redefinition of
$\mu$~\eqref{muWL} as
\beq
2 a^2  \sum_{i=1}^n  \left(- {x_{i i+2}^2\over  \mu^2}\right)^{- 2 \e}
\,
\Big[{\Gamma ( 1 + \e) e^{  \gamma \e}}\Big]^2 \, f_\mathrm{cusp} (\e )
\ .
\eeq
Since
\beq
\Big[{\Gamma ( 1 + \e) e^{  \gamma \e}}\Big]^2 \, f_\mathrm{cusp} (\e ) \ = \
{\pi^2\over 48 \e^2} \,  -\,  {\zeta_3\over 8 \e} \  + \   {7 \pi^4 \over 720}\ + \   \cO (\e)
\ ,
\eeq
we obtain that  the contribution from all cusps is therefore
\beq
\label{cuspours}
2 a^2\left[  \, \sum_{i=1}^{6} \left( - {x_{i i+2}^2\over \mu^2} \right )^{-2 \e}  \Big( {\pi^2\over 48 \e^2} \,  -\,  {\zeta_3\over 8 \e} \Big)
\ + \  {7 \pi^4 \over 120} \right]
 \ + \   \cO (\e)
\ .
\eeq
Equation \eqref{cuspours} is exactly equal to  the  corresponding cusp results
in \cite{dhks6} (after considering that their $\e$ is  the ultraviolet parameter).

\section{Evaluation of  Wilson loop diagrams}

An intriguing property of the
$n$-point polygon Wilson loop parameterisations from Section~\ref{sec:wilsondiags} is that they are valid
for  an arbitrary $n$.
The infrared properties of the integrals needed to evaluate the diagrams
may change from more to less divergent,
however all cases  can be  obtained from the same starting
expressions involving parametric integrals with five dimensions at most.
This is to be contrasted with the Feynman parameterisations  of
Feynman amplitudes, which require a  different representation with
additional parameters as the number of external legs $n$ is increased.

Our goal is  to construct an algorithm which evaluates the diagrams of the
previous section in complete generality, for  an arbitrary number of
points, $n$.  This has been achieved for up to  $n=6$ in the
literature \cite{dhks}.
Given the  fact that the number of Feynman parameter integrations and
the number of distinct ``master'' functions required for an evaluation
at different values of $n$ is independent of it, one could aim for solving this  problem
in complete generality.   It is very exciting that Wilson loops and
$\cN =4$ SYM  planar MHV amplitudes are very likely to be dual to each other.
If this   is proven correct, solving the  problem of  calculating two-loop $n$-gon Wilson loops
also provides a solution to the problem of evaluating  planar two-loop amplitudes with an
arbitrary number of legs.

A fully analytic evaluation of the  master functions in
arbitrary $n$-point Wilson loops appears to be
a formidable task. A more viable and practical approach  is
to evaluate these  integrals numerically.
Many of  the  required integrals  develop divergences  when $\epsilon =0$,
with  $1/\epsilon^2$ poles at most.  We use the programs developed
in~\cite{Anastasiou:2008rm, Anastasiou:2007qb, Lazopoulos:2007ix, Anastasiou:2005pn}
for evaluating  generic Feynman diagrams
in order to automatically cast these integrals as  Laurent series in $\epsilon$.
The coefficients of the series are multidimensional integrals, however
they are free of  singularities in the dimensional regularisation parameter,
and  we can evaluate them numerically using well established
stochastic  integration methods~\cite{Hahn:2004fe}.

We notice that  we  have evaluated  the  finite  ``hard  diagrams''  (which appear for the first time at
$n \geq 6$ points),
using an alternative approach; this is possible because the
diagram is free of singularities in the limit $\epsilon \to 0$.
We  recall that a one-loop scalar triangle can be  written using Feynman
parameters as
\begin{eqnarray}
\label{eq:triangle}
&&
{\rm Tria}(D; \nu_1, \nu_2, \nu_3) :=  \int \frac{d^Dk}{i\pi^{\frac{D}{2}}}
\frac{1}{
\left( k^2 \right)^{\nu_1}
\left[ (k+p_1)^2 \right]^{\nu_2}
\left[ (k+p_1+p_2)^2 \right]^{\nu_3}
}   \\
&=&(-1)^{\nu_{123}} \frac{\Gamma\left(\nu_{123}-\frac{D}{2} \right)}
{\Gamma(\nu_1)\Gamma(\nu_2)\Gamma(\nu_3)}
\int\!\!\prod_{i=1}^{3} d \a_i \
\delta(1 - \sum_{i=1}^3 \a_i )
{
\a_1^{\nu_1-1} \a_2^{\nu_2-1} \a_3^{\nu_3-1}  \over
(\a_1 \a_2 p_1^2 + \a_2 \a_3 p_2^2 + \a_3 \a_1 p_3^2 )^{\nu_{123}   -\frac{D}{2}}
},  \nonumber
\end{eqnarray}
with $p_3=-p_1-p_2$ and $\nu_{123} := \nu_1 + \nu_2 + \nu_3$.
Interestingly, this is the same  function as the triple-gluon vertex of
\eqref{G}. In the absence of divergences,
as occurs in the hard diagrams with nonvanishing $Q_1$, $Q_2$ and $Q_3$,
we can set $D=4$ (equivalently $\epsilon=0$) in
\eqref{G}. By comparing \eqref{G} and \eqref{eq:triangle},
we observe that the triple-gluon vertex is  a sum of one-loop scalar triangles in $D=6$ dimensions,
with powers  of propagators taking the values $\nu_i=1,2$.

We exploit the mapping of the vertex of \eqref{G} onto familiar one-loop integrals to simplify it before we insert its expression into the hard diagram. Using an automated reduction program~\cite{Anastasiou:2004vj},
we express the finite part  of the vertex as  a linear combination
of  triangle  ${\rm Tria}(6-2\epsilon; 1, 1, 1)$, and bubble
${\rm Tria}(4-2\epsilon; 0, 1, 1)$, ${\rm Tria}(4-2\epsilon; 1, 0, 1)$,
${\rm Tria}(4-2\epsilon; 1, 1, 0)$ master integrals.
Notice that we have re-introduced the dimensional  regulator
$\epsilon$ because these master integrals are all divergent as $\epsilon \to 0$.
However, we can set $\epsilon=0$ after  we substitute the analytic expressions
for the bubble  master integrals, and a Feynman representation derived
from \eqref{eq:triangle} for the  triangle master integral in six
dimensions. An alternative basis of master integrals can  be
obtained  using  dimensional shift identities~\cite{Davydychev:1991va,Anastasiou:1999bn},
where the triangle master integral is also in four dimensions.
However, this choice is inconvenient
for numerical evaluations,  because the dimensional shift generates a
Gram determinant in the denominator
$$
\frac{1}{\hat{\Delta}_3(p_1^2, p_2^2, p_3^2)},
$$
which spoils the numerical convergence  of  stochastic integration.

The calculation of the ``hard diagram'' is  performed with a five-dimensional
numerical integration after expressing the vertex in terms of master integrals.
A na\"{i}ve numerical integration of the expression in \eqref{eq:hardexpression},
without a  reduction of the vertex to master integrals,
is also five dimensional.
However, our reduction method has the advantage of
removing integrable singularities which may emerge
in certain kinematic limits. Importantly, the na\"{i}ve numerical
integration method becomes  unstable for $n \geq 8$, while the combination
of  the reduction with numerical integration is stable and  efficient.

As we have  mentioned, the number of  ``master'' functions required is  independent
of  $n$.  However, there  are many possibilities for the kinematic invariants
which enter as  arguments of  these functions  -- they can be  squares  of either lightlike
or  massive momenta.
However, we notice that evaluating the same master integrals with some of the kinematical invariants equal to
zero  may (and often does) yield a different structure of infrared or  integrable singularities.
Therefore,  these  cases  are treated distinctly  in our numerical approach.

In the rest of the paper, we  present explicit results
for $n=6,7,8$. This  exhausts all possibilities for the distinct
configurations in the  evaluation of the diagrams of the previous session.
The computation of  $n > 7$ polygon Wilson loops  proceeds with an identical algorithm
as  for $n=7$.

\section{Six-point Wilson loops  }

An analytic form of the six-point remainder function $\cR^\mathrm{WL}_6(u_1,u_2,u_3)$, where the six-point cross-ratios $u_{1,2,3}$ have been defined in \eqref{sixptcr}, is currently not available. However we have used our numerical methods to map this function in a number of ways, as we will now report.

Before doing this, we note that some numerical calculations of values of the six-point remainder function
were presented in \cite{dhks6}. As an initial check on the validity and consistency of our methods, we compared our results with those of this reference; we found complete agreement.
Recall that we are using a different definition for the remainder function than \cite{dhksbum,dhks6},
which for the hexagon Wilson loop implies $\cR^\mathrm{WL}_6(u_1,u_2,u_3) = \cR^\mathrm{DHKS}_6(u_1,u_2,u_3)-c_\mathrm{W}$.
In \cite{dhks6} $c_\mathrm{W}$ was found to be equal to $12.1756$ with an absolute accuracy of
about $10^{-3}$ and we observe that this is close to $\pi^4/8 \sim 12.1761$.
For example, studying the collinear limit with our numerical routines we found
\beq\label{collinear6pt}
\cR^\mathrm{WL}_6(0,u,1-u) = 0 \pm 0.01\, ,
\eeq
in agreement with \cite{dhksbum}. Note that one cannot simply calculate values of this
function with one of the $u$ variables set to zero, as the errors grow as any of the
variables approaches zero; however one can plot the functions obtained for
$\cR^\mathrm{WL}_6(u_1,u,1-u)$ for various values of $u_1$ and see that this function becomes flat as $u_1\rightarrow 0$ \cite{dhksbum}. We have done this with our routines and find the value $0\pm0.01$,
in agreement with the value $12.1756$ found in \cite{dhks6} for $\cR_6^\mathrm{DHKS}$ in the collinear limit. Furthermore,  \eqref{collinear6pt} implies that $\cR_6^\mathrm{WL} (0,0,1) = 0$, which is consistent with the predictions of
\cite{Bartels:2008ce,Bartels:2008sc,Brower:2008ia,DelDuca:2008jg}
derived in the multi-Regge kinematics (at least in the case where all the kinematic invariants are defined
in the Euclidean region $-s \gg -s_i \gg -t_i > 0$).

In Table 2 of \cite{dhks6}, a number of values of $\cR_6^\mathrm{DHKS}$ are also listed for different kinematics.
We have checked the values of the remainder function for all these inputs and are in perfect
agreement with the quoted results up to the ubiquitous constant $c_\mathrm{W}$.

The remainder function $\cR^\mathrm{WL}_6(u_1,u_2,u_3)$ is also symmetric under permutations of the three cross-ratios. We have checked this in various particular cases and it is also
amply confirmed by the results plotted in the graphs below. Before we discuss some plots of our
numerical results in more detail
we wish to add a couple of comments on the dual conformal invariance  of $\cR^\mathrm{WL}_6(u_1,u_2,u_3)$, which, if correct, implies
that  in the six-point case the remainder function should depend on the explicit gluon momenta only through the cross-ratios
$u_1, u_2, u_3$. We have confirmed this expectation by various numerical tests
where we held the cross-ratios fixed but varied the Mandelstam variables $x^2_{ii+2}$ and $x^2_{ii+3}$.
Furthermore, we have tested conformal invariance for kinematic points that obey the Gram determinant
constraints (strictly four-dimensional kinematics) and for kinematic points that do not obey the Gram
determinant constraints. We always found perfect agreement within our numeric accuracy for
$\cR^\mathrm{WL}_6(u_1,u_2,u_3)$ as long as the cross-ratios were held fixed.

\begin{table}[h]
\begin{center}
\begin{tabular}{|c||c|c|c|}
\hline $(u_1,u_2,u_3)$ &
 $\cR^\mathrm{WL}_6(A)$ &$\cR^\mathrm{WL}_6(B)$ & $\cR^\mathrm{WL}_6(C)$
\\
\hline
\hline  $(1/9,1/9,1/9)$    & 5.18056  & 5.18096 & 5.18102
\\
\hline   $(1/4,1/4,1/4)$   & 1.08916  & 1.08916 & 1.08919
\\
\hline $(1,1,1)$               & -2.70814 & -2.7066& -2.70657
\\
\hline $(100,100,100)$ & -2.09134 & -2.09204& -2.09228
\\
\hline
\end{tabular}
\end{center}
\caption{\it Checks of conformal invariance of the remainder function $\cR^\mathrm{WL}_6$. In each horizontal line we present values of $\cR^\mathrm{WL}_6$ for different kinematic points (A), (B) and (C) that yield the same  cross-ratios $(u_1,u_2,u_3)$. We find that within our numerical, absolute errors $\pm 0.01$ the values match perfectly. Note that this estimate of the errors is rather conservative and that the actual error is closer to $\pm 0.001$.}
\label{confinvtable}
\end{table}

In the following we give a couple of explicit examples to demonstrate dual conformal invariance
of the remainder function at six points.
To be more concrete we consider
four kinematic points $u_1=u_2=u_3=1/9 , \, 1/4, \, 1, \, 100$, with (A) $x^2_{ii+2}=-1$, (B) $x^2_{13}=x^2_{24}=-1 , \, x^2_{35}=x^2_{46}=x^2_{51}=x^2_{62}=-2$ and
(C) $x^2_{13}=x^2_{24}=-1 , \, x^2_{35}=x^2_{46}=-2 , \, x^2_{51}=x^2_{62}=-3$. The numerical results
for these kinematic points are collected in Table \ref{confinvtable}.
In general these kinematic configurations do not obey the Gram determinant constraint, but we have checked for  numerous values of $(u_1, u_2, u_3)$ that $\cR^\mathrm{WL}_6$ is
independent of the Gram determinant constraint as was also observed in \cite{dhks6} for one particular
set of cross-ratios. We will here only discuss in detail the case $u_1=u_2=u_3=100$ for which one
possible kinematic point, consistent with the Gram determinant constraint, is given by:
\beqa
&&x^2_{13}=x^2_{35}=x^2_{46}=x^2_{62}=-\frac{20+3 \sqrt{42}}{2} \, , \,\, x^2_{24}=x^2_{51}=-1 \, ,\nonumber \\
&&x^2_{14}=x^2_{25}=-\frac{1}{10} \, , \,\, x^2_{36}= -\frac{389+60 \sqrt{42}}{20} \, .
\eeqa
For this particular kinematic configuration the numerical evaluation of the remainder functions yields
$-2.099$ with an absolute error of about $\pm 0.02$, which is in agreement with the last row of Table \ref{confinvtable}.

\begin{figure}[ht]
\begin{center}
\includegraphics[width=12cm]{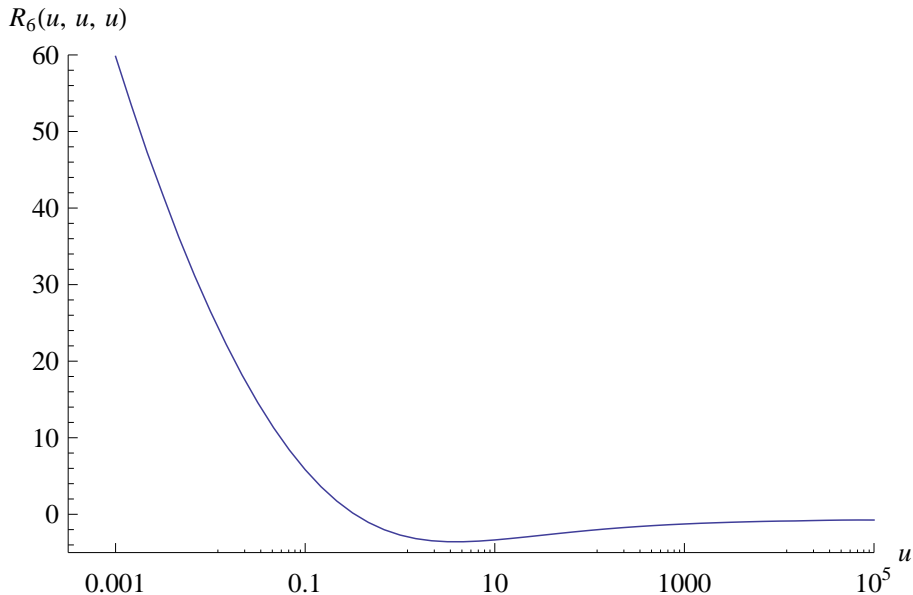}
\end{center}
\caption{\it A plot of the remainder function of the hexagon Wilson loop with $u_{1}=u_{2}=u_{3}=u$.
For the location of the minimum of this function we found a numerical value of $u=3.83\pm0.01$.}
\label{diagRsix}
\end{figure}
We now turn to describe additional numerical results we have found
for $\cR^\mathrm{WL}_6(u_1,u_2,u_3)$. In order to explore possible
analytic expressions for this function, we first considered \beq
F_6(u) \equiv \cR^\mathrm{WL}_6(u,u,u). \eeq A plot of this
function is given  in Figure \ref{diagRsix}. Salient features are
the minimum value of $F_6(u)$ which is $-3.60(\pm0.01)$ at $u=
3.83(\pm 0.01)$ and the asymptotic value $F_6(u)\rightarrow
-0.67(\pm0.05)$ as $u\rightarrow \infty$. Another special value is
$F_6(1)=-2.706(\pm 0.007)$. It is interesting to observe that the
minimum of $F_6(u)$ and $F_6(1)$ are well approximated by
transcendentality four numbers, namely $-\frac{\pi^4}{27} \sim
-3.6077$ and $-\frac{\pi^4}{36} \sim -2.7058$, respectively.

Some
of the generic features of $F_6(u)$ are reproduced well by
transcendentality four functions.
For example, we comment that ${\rm ln} (u)\,{\rm Li}_3(-1/u) + (1/3)\, {\rm ln}(u)^2 \, {\rm Li}_2(-1/u)$ matches
the asymptotic behaviour of $F_6(u)$, behaving as
${ const} \cdot {\rm ln}(u)^2$ at $u\rightarrow 0$ and going to a constant as $u\rightarrow \infty$, as well as
matching its general shape. Whilst
this is perhaps encouraging, at present we do not have a global
match of our numerical results to an explicit function with
transcendentality four. We also observe from our data that $F_6(u)
\sim {\ln}^2 (u)$ for asymptotically small values of $u$. If the
multiplicative constant is equal to rational number times $\pi^2$,
this is consistent with transcendentality four behavior.

\begin{figure}[ht]
\begin{center}
\includegraphics[width=12cm]{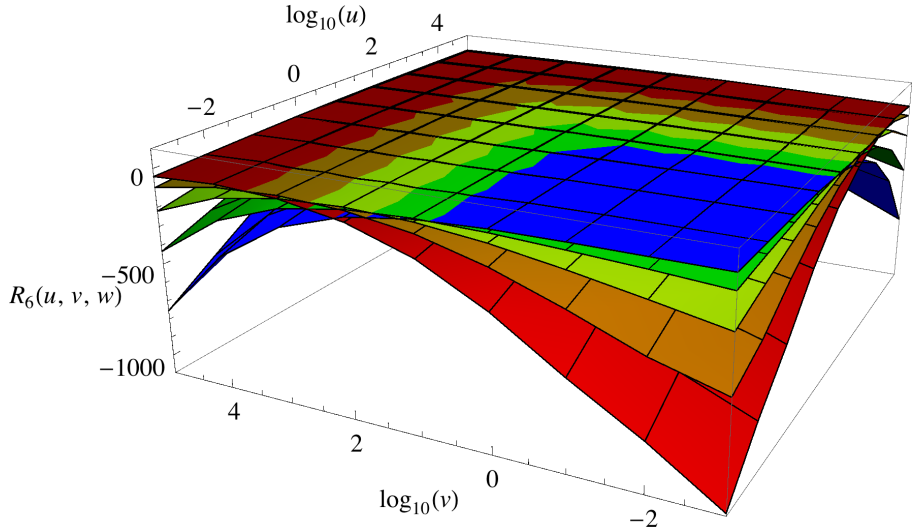}
\end{center}
\caption{\it This graph contains five plots of the remainder function of the hexagon Wilson loop with $u_{1}=u$, $u_{2}=v$ and $u_{3}=w$.
The cross-ratios $u$ and $v$ vary between $10^{-3}$ and $10^5$ while for $w$
we have chosen five fixed values: $w=1$ blue plot, $w=10$ green plot, $w=100$ yellow plot, $w=1000$ orange plot,
and $w=10000$ red plot.}
\label{Rsix12345v1}
\end{figure}

\begin{figure}[ht]
\begin{center}
\includegraphics[width=12cm]{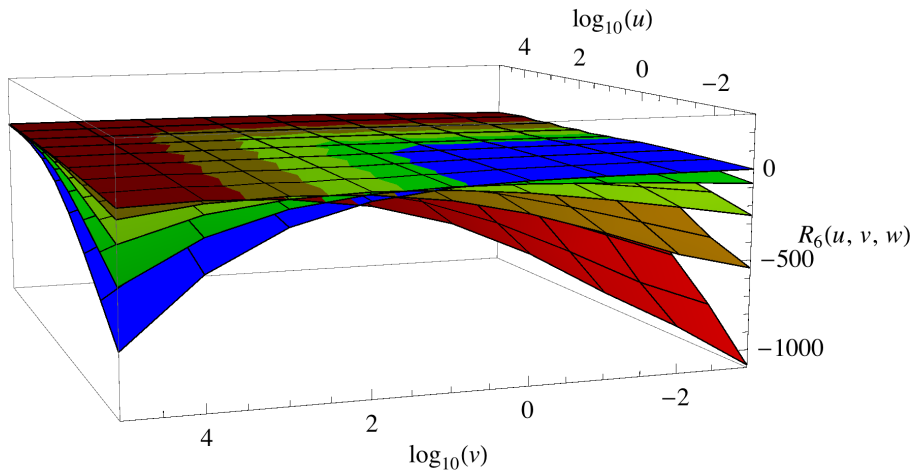}
\end{center}
\caption{\it This graph contains five plots of the remainder function of the hexagon Wilson loop with $u_{1}=u$, $u_{2}=v$ and $u_{3}=w$. The cross-ratios $u$ and $v$ vary between $10^{-3}$ and $10^5$ while for $w$
we have chosen five fixed values: $w=1$ blue plot, $w=10$ green plot, $w=100$ yellow plot, $w=1000$ orange plot,
and $w=10000$ red plot.}
\label{Rsix12345v2}
\end{figure}

A more complete picture of the structure of the remainder function may be obtained by
exhibiting a selection of slices of $\cR^\mathrm{WL}_6(u,v,w)$ at different values of $w$. Due to the slowly changing behaviour of the function as $u$ and $v$ vary, it proves instructive to give log-based plots -- in the following the $u, v$ coordinates run from $10^{-3}$ to
$10^{5}$ (listed as $-3,...,5$ in the Figure). In Figure \ref{Rsix12345v1} and Figure \ref{Rsix12345v2} we present from different viewpoints plots of the function $\cR^\mathrm{WL}_6(u,v,w)$ for these values of $(u,v)$, in five cases where $w=1, 10, 100, 1000, 10000$.

One may make a number of comments regarding these plots. Firstly, the symmetry of the function under the interchange of $u$ and $v$ is apparent (and is manifest in the actual data). Secondly, for the values of $w$ considered, for small $u, v$
the remainder function takes a large negative value for $w$ large, and increases as $w$ decreases. The order of these hyper-surfaces reverses
as $u$ or $v$ increase, and for large values of these variables, the remainder function becomes increasingly negative (as of course required by the symmetry property and the behaviour at large $u$). For all three variables large, the remainder function approaches a constant which is equal to the
asymptotic value of $F_6(u)$ (about $-0.67$). In general, it is apparent from Figures \ref{diagRsix}-\ref{Rsix12345v2} that the remainder function is rather smooth for all values of the cross-ratios.


\section{Seven-point Wilson loops and collinear limits}

In this section we wish to address two separate  issues.

 Firstly, we present numerical evidence that the
seven-point Wilson loop remainder function is a function of the appropriate seven-point
cross-ratios.
As anticipated in   Section~\ref{sec:remainder-function},
we define these cross-ratios without requiring the Gram determinant
conditions, therefore we expect to have seven cross-ratios at seven points.
As a basis of seven independent cross-ratios at seven points, we will choose the following
quantities,
\begin{equation}
\label{7pntbasis}
  u_{14},\ u_{25},\ u_{36},\ u_{47},\ u_{15},\ u_{26},\ u_{37}
  \ ,
\end{equation}
where $u_{ij}$ is defined in \eqref{ourcrossratios}.

Next, we  will study how the remainder function behaves under collinear limits.
In particular, we will see that the seven cross-ratios in \eqref{7pntbasis} will naturally flow into
four parameters, three of which are naturally related to the three six-point cross-ratios;
we will then present evidence that there is in fact no dependence on  the fourth
parameter, related to the parameter $z$ introduced in the collinear limit (see \eqref{abcoll} below).
As we shall discuss,  our results support the
conjecture that the remainder function of the Wilson loops should be equal
to the corresponding remainder function on the amplitude side.

\subsection{Seven-point remainder function and conformal invariance}

We can now compute two-loop contributions to the logarithm of the seven-point
Wilson loop for arbitrary kinematics. There are fourteen kinematic variables
formed by seven two-particle invariants and seven three-particle invariants.
In this case we keep the two-particle invariants as independent inputs, and
trade the three-particle invariants for the seven conformally
invariant cross-ratios defined above.

The sum of all relevant diagrams gives rise to the
two-loop contribution to the logarithm of the Wilson loop. After
subtracting from it the known BDS expression we find
the  remainder function $\cR_7^\mathrm{WL}$.
It follows from our numerical calculations that $\cR_7^\mathrm{WL}$ is
independent of the
non-conformal input
(in this case the seven two-particle invariants)
and is only a function of the cross-ratios:
\beq
\label{Rseven}
\cR_7^\mathrm{WL} \, =\, \cR_7^\mathrm{WL}(
u_{14},u_{25},u_{36},u_{47},u_{15},u_{26},u_{37})
\ .
\eeq
This function is also invariant under cyclic permutations of all $u$'s
and under the reflection symmetry (which exchanges the clockwise with the
anticlockwise ordering of $u$'s).

Below we give some explicit examples to demonstrate dual
conformal invariance
of the remainder function at seven points and invariance under cyclic
permutations and
reflection.
To be more concrete we consider kinematic points for various values of the
conformal cross-ratios with (A) $x^2_{ii+2}= -1 $ and
(B) $x^2_{ii+2}=-i$ for $i=1 \ldots 7$. The numerical results
for these kinematic points are collected in Table \ref{table7point}.
\begin{table}[h]
\begin{center}
\begin{tabular}{|c||c|c|}
\hline
$(u_{14},u_{25},u_{36},u_{47},u_{15},u_{26},u_{37})$ & $\cR^\mathrm{WL}_7(A)$ &
$\cR^\mathrm{WL}_7(B)$  \\
\hline
\hline
$(1,1,1,1,1,1,1)$  & -3.85627  & -3.85732 \\
\hline
$(1/4,1/4,1/4,1/4,1/4,1/4,1/4)$ & 8.13063 & 8.13272 \\
\hline
$(1/4,1,1,1/4,1,1,1)$ & -4.40748 & -4.40651 \\
\hline
$(1,1/4,1,1,1/4,1,1)$ & -4.40657 & -4.40056 \\
\hline
$(1,1,1/4,1,1,1/4,1)$    & -4.40654  & -4.40559 \\
\hline
$(1,1,1,1/4,1,1,1/4)$     & -4.40746 & -4.40617 \\
\hline
$(1,1/2,1,1,1,1/4,1)$      & -4.27219 & -4.27108 \\
\hline
$(1,1/4,1,1,1,1/2,1)$   & -4.27224 & -4.27049 \\
\hline
$(1/4,1,1/4,1,1,1,1)$ & -4.63668 & -4.63696 \\
\hline
\end{tabular}
\end{center}
\caption{\it Checks of conformal invariance and invariance under
cyclic permutations of the $u$'s and
reflection symmetry of the remainder function $\cR^\mathrm{WL}_7$. In
each horizontal line we present values of $\cR^\mathrm{WL}_7$ for
different kinematic points (A) and (B)  that yield the same  cross
ratios. We find that within our numerical, absolute errors, which
range between $\pm0.001$ and $\pm 0.01$ for individual kinematic
configurations, the values match nicely.}
\label{table7point}
\end{table}

\begin{figure}[h]
\begin{center}
\includegraphics[width=12cm]{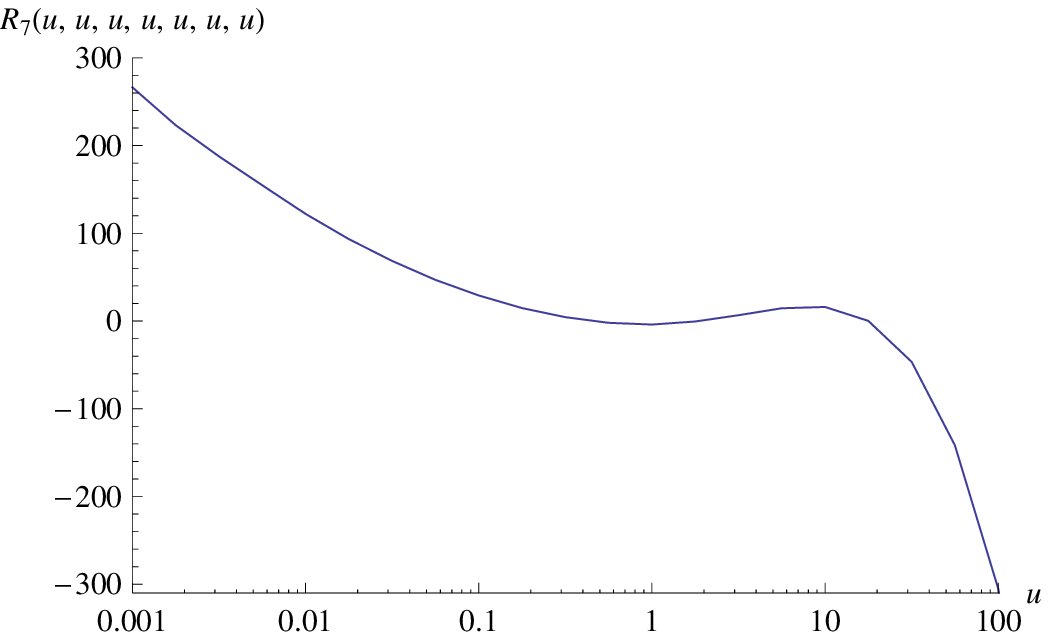}
\end{center}
\caption{\it A plot of the remainder function of the seven-sided
  Wilson loop with all cross-ratios equal to $u$. }
\label{fig:7ptu}
\end{figure}

We have computed the remainder function for many other values of the
cross-ratios.
In Figure~\ref{fig:7ptu} we display the remainder function when all
cross-ratios are equal.

\subsection{Simple collinear limits}
\label{collinearsection}

In this section we present numerical evidence that collinear limits of $n$-gon Wilson loops
with $n=7$  behave  in the same way as the corresponding amplitude collinear limits. Collinear limits have been used
very often as tools to check the consistency
of ansatze for the expression of infinite sequences of scattering amplitudes,
see for example \cite{bddk}.
In much the same way, by showing that Wilson loops, and in particular the remainder function, have the
expected collinear limits, we can provide further evidence in support of the conjectured duality between
the finite parts of these two a priori completely different quantities.

We begin with a brief review of the universal behaviour of scattering amplitudes under simple collinear limits.
In this limit, one  selects two adjacent momenta
$p_a$ and $p_{b}$, and sets
\beq
\label{abcoll}
p_a \ = \ z P\ , \qquad \qquad
p_{b}\  = \  (1-z) P
\ .
\eeq
The collinear limit is taken by
letting $P^2 \to 0$.  Under this limit, scattering amplitudes behave in a well-known, universal way.
Consider for instance a one-loop scattering amplitude,
$\cA_n^{1-{\rm loop}}$. When the two momenta $p_a$ and $p_b$ become collinear,
the amplitude is known to factorise as \cite{bddk,fusing,Bern:1995ix,Kosower:1999xi}
\beqa
\label{coll}
&& \cA_n^{1-{\rm loop}}
(1,\ldots, a^{\l_a}, b^{\l_b}, \ldots, n)
\;
{\buildrel a \parallel b\over
{\relbar\mskip-1mu\joinrel\longrightarrow}}
\\ \nonumber
&& \hspace{1 cm}\sum_{\s} \bigg[
 {\rm Split}^{\rm tree}_{- \s} ( a^{\l_a}, b^{\l_b})
\ \cA_{n-1}^{1-{\rm loop}}
(1,\ldots, (a+ b)^{\s},  \ldots, n)
\nonumber \\
&& \hspace {1.4cm}
\ + \
{\rm Split}^{1-{\rm loop}}_{- \s} ( a^{\l_a}, b^{\l_b})
\ \cA_{n-1}^{\rm tree}
(1,\ldots, (a+ b)^{\s},  \ldots, n)
\bigg]
\ .
\nonumber
\eeqa
${\rm Split}^{\rm tree}$ are  tree-level
splitting amplitudes, whose explicit forms can be found,
for instance,  in  \cite{Dixon:1996wi}.
${\rm Split}^{1-{\rm loop}}$ is a
one-loop splitting amplitude.
Explicit formulae for this one-loop splitting
amplitude, valid to all orders
in the dimensional regularisation
parameter $\e$, were presented in \cite{ku} and \cite{vittorio}.
We quote here the result of \cite{vittorio} for
the $\cN=4$ theory:
\beq
\label{susysplit}
{\rm Split}^{1-{\rm loop}}_{- \s} ( a^{\l_a}, b^{\l_b})
\ = \
{\rm Split}^{\rm tree}_{- \s} ( a^{\l_a}, b^{\l_b})
\ r_S^{(1)} (\e; z, s_{ab} )
\ ,
\eeq
where, to all orders in $\e$,
\beq
\label{cl}
r_S^{(1)} (\e; z, s_{ab} ) \ := \
{\hat{c}_\Gamma \over \e^2} \Big( {-s_{ab} \over \mu^2} \Big)^{-\e}
\left[  1 \, - \,
\mbox{}_{2}F_1 \left( 1, -\e, 1- \e, {z-1 \over z}\right)
  \, - \, \mbox{}_{2}F_1 \left(1, -\e, 1- \e, {z \over z-1}\right)
\right]
\ ,
\eeq
and $\hat{c}_\G$ is defined in~(\ref{cg}).

We now move on to consider the
behaviour of the remainder function $\cR_n$ as defined in \eqref{remainder} under collinear limits, following
\cite{seven}.
Consider then the two-loop term $\cM^{(2)}_n(\e )$ in the expansion of the
amplitude, and write it as
\beq
\label{fm}
\cM^{(2)}_n(\e ) - {1\over 2} \Big( \cM^{(1)}_n (\e) \Big)^2 \ = \ f^{(2)} (\e) \cM^{(1)}_n  ( 2 \e ) \, + \,  C^{(2)} \, + \, \cR_n \
\, + \, \cO (\e )
\ ,
\eeq
where $f^{(2)} (\e):= f^{(2)}_0 + f^{(2)}_1 \e + f^{(2)}_2 \e^2$ and $ C^{(2)}$ are defined in \eqref{f2}, \eqref{c2}.
Using \eqref{coll} and \eqref{susysplit}, one  sees that, under a simple
collinear limit,  the scalar function $\cM^{(1)}$ must behave as   \cite{bds}
\beqa
\label{prima}
\cM^{(1)}_n & \to  &  \cM^{(1)}_{n-1} + r_S^{(1)} (\e; z, s_{ab}) \ ,
\\ \nonumber
\cM^{(2)}_n & \to  &  \cM^{(2)}_{n-1}  + r^{(1)}_S (\e; z, s_{ab} ) \cM^{(1)}_{n-1} + r_S^{(2)} (\e; z, s_{ab} )
\ .
\eeqa
It was shown in  \cite{abdk} that splitting amplitudes obey an iterative formula identical to the
homogeneous form of the
BDS conjecture for the amplitude,  i.e.
\beq
\label{babis2}
r^{(2)}_S(\e; z, s_{ab})  - {1\over 2} \big( r^{(1)}(\e; z, s_{ab})  \big)^2 \, = \, f^{(2)} (\e) r_S^{(1)} ( 2 \e; z, s_{ab}  ) \, + \cO (\e )
\ .
\eeq
Using \eqref{prima} and \eqref{babis2}, one  sees that, under a simple collinear limit,
\begin{align}
\cM^{(2)}_n(\e ) &- {1\over 2} \Big( \cM^{(1)}_n (\e) \Big)^2 - f^{(2)}(\e)  \cM_n^{(1)} ( 2 \e ) \nonumber \\
 &\rightarrow
\cM^{(2)}_{n-1}(\e ) - {1\over 2} \Big( \cM^{(1)}_{n-1} (\e) \Big)^2 - f^{(2)}(\e)  \cM_{n-1}^{(1)} ( 2 \e )
\ .
\label{vh}
\end{align}
Equation \eqref{fm}
defined the finite remainder function $\cR_n$ of the amplitude as
\beq
\label{fm2bis}
\cR_n \ = \    \cM^{(2)}_n(\e )\, - \, {1\over 2} \Big( \cM^{(1)}_n (\e) \Big)^2 -  f^{(2)} (\e) \cM^{(1)}_n  ( 2 \e ) \, - \,  C^{(2)}
\, + \cO (\e ) \
\ ,
\eeq
and it follows from  \eqref{vh} that in the simple collinear limit $\cR_n \   \to \   \cR_{n-1}$, as anticipated in
\eqref{abbc}.

What about simple collinear limits of Wilson loops? If the duality with amplitudes holds, we expect that
the Wilson loop will have the same collinear limits as the amplitude, as discussed in  \eqref{abb}.

Let us now specify this discussion to the seven-point Wilson loop case.
Specifically, in the simple collinear limit of a seven-point amplitude one expects to find
\beq
\label{Rsevencollgen}
\cR_7^\mathrm{WL} \left(u_{14},u_{25},u_{36},u_{47},u_{15},u_{26},u_{37} \right) \rightarrow \cR_6^\mathrm{WL} (u_1,u_2,u_3) \ ,
\eeq
where the seven-point basis of cross-ratios is defined in \eqref{7pntbasis}, while the six-point basis in
\eqref{sixptcr}. Notice that on the right hand side of \eqref{Rsevencollgen} we do not allow for an additional constant term.

For concreteness we take $p_6$ and $p_7$ to be collinear. We set
\beq
\label{coll67}
p_6 = x_6-x_7 = z P \  , \qquad p_7 = x_7-x_1 = (1-z) P \ ,
\eeq
where as usual $P^2 \to 0$ in the collinear limit. In this collinear limit, the cusp at $x_7$ is therefore
``flattened".
This collinear limit of the seven-point kinematics is characterised by
\bea
&&x^2_{16} \to 0 \ , \qquad x^2_{27} = (1-z)x^2_{26} \ , \qquad  x^2_{57} = z x^2_{51} \ , \\
&& x^2_{37} = (1-z)x^2_{36} +  z x^2_{13} \ , \qquad x^2_{47} = (1-z)x^2_{46} +  z x^2_{14} \ , \nonumber\
\eea
with all other $x^2_{ij}$ segments unchanged.

\begin{figure}[t]
\begin{center}
\scalebox{1.30}{
\includegraphics[width=12cm]{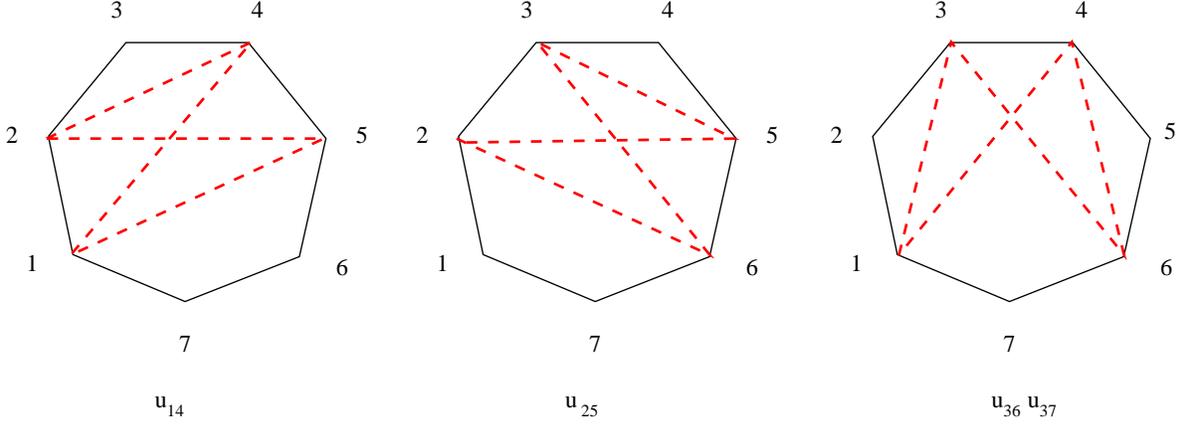}
}
\end{center}
\caption{\it  The cross-ratios $u_{14}$, $u_{25}$ and $u_{36}u_{37} $ of
the heptagon Wilson loop which do not include the cusp at $x_7$.
The red dashed lines depict $x_{ij}^2$ segments on the right hand sides of
\eqref{u67717}.}
\label{fig:u671}
\end{figure}
As one can see in Figure \ref{fig:u671},
there are three cross-ratios which do not pass through
the cusp at  $x_7$. The first two are members of our
basis \eqref{7pntbasis},  whereas the third one
is a product of members of our basis,
\beq
\label{u67717}
u_{14} :=  {x_{15}^2 x_{34}^2 \over x_{14}^2 x_{35}^2}
\ , \qquad
u_{25} :=  {x_{26}^2 x_{35}^2 \over x_{25}^2 x_{36}^2}
\ , \qquad
u_{36}\,u_{37}  :=  {x_{13}^2 x_{46}^2 \over x_{14}^2 x_{36}^2}
\ .
\eeq
These ratios agree precisely with the three variables of the six-point case,
\beq
\label{u7to6}
u^{(7)}_{14} \rightarrow u^{(6)}_{14}=u_2 \ , \qquad u^{(7)}_{25}
\rightarrow u^{(6)}_{25}=u_3
\  ,
\qquad  u^{(7)}_{36} \,u^{(7)}_{37} \rightarrow u^{(6)}_{36}=u_1 \ .
\eeq
where the superscripts on the $u$'s denote the number of edges of the
corresponding Wilson loop and $u_1,u_2,u_3$ are the six-point
cross-ratios defined in \cite{dhksbum} and given in \eqref{sixptcr}.
Specifically, the seven-point cross-ratios defined in \eqref{7pntbasis} become
(see \eqref{ourcrossratios})
\bea
\label{useven2col}
\nonumber
&&u_{14} = {x_{24}^2 x_{15}^2 \over x_{25}^2 x_{14}^2}  \ , \quad
u_{25} = {x_{26}^2 x_{35}^2 \over x_{25}^2 x_{36}^2} \ ,\\
&&u_{36} = {x_{46}^2 \over x_{36}^2}{z x_{13}^2 + (1-z)x_{36}^2 \over z x_{14}^2 + (1-z)x_{46}^2}  \ , \quad
u_{47} = {z x_{14}^2 \over z x_{14}^2 + (1-z)x_{46}^2}  \ , \quad
u_{15} =  0 \ ,  \nonumber \\
&&u_{26}= {(1-z)x_{36}^2 \over z x_{13}^2 + (1-z)x_{36}^2}  \ ,
\quad u_{37} = {x_{13}^2 \over x_{14}^2}{z x_{14}^2 + (1-z)x_{46}^2 \over z x_{13}^2 + (1-z)x_{36}^2}
 \ .
\eea
As expected, it follows immediately
that $u_{14}$, $u_{25}$ and $ u_{36}\, u_{37} $ are
equal to the three cross-ratios of the six-point case, as dictated by \eqref{u7to6}.
In addition, \eqref{useven2col} imply that there are three constraints on the remaining four variables,
\beq
\label{constrs2}
u_{15} =  0 \ , \quad
u_{47} + u_{26} \,u_{36} =1 \ , \quad  u_{26} + u_{37}\,u_{47} =1 \ .
\eeq
Taking this into account, and solving the two constraints in \eqref{constrs2} for $u_{47}$ and $u_{26}$ we
conclude that the collinear limit relates the seven-point remainder function to the six-point one
as follows:
\begin{align}
  \label{Rsevencoll}
\cR_7^\mathrm{WL}( u_{14},u_{25},u_{36},u_{47},u_{15},u_{26},u_{37}) &\rightarrow
  \cR_7^\mathrm{WL}\left(u_{14},u_{25},u_{36},\frac{1-u_{36}}{1-u_{37}u_{36}},0,
    \frac{1-u_{37}}{1-u_{37}u_{36}},u_{37}\right) \nonumber\\
&= \cR_6^\mathrm{WL}  (u_{37}u_{36},u_{14},u_{25}) \ .
\end{align}
%
%
Note that three of the seven variables on the left hand side of \eqref{Rsevencoll} are constrained, leaving four
variables free. Since the right hand side is a function of only three variables, it follows that the left hand side
actually does not depend on one combination:
$u_{37}/u_{36} :=\kappa$.
Thus we can rewrite \eqref{Rsevencoll} in terms of the six-point
cross-ratios $u_i$ and $\kappa$ as
\beq
\label{Rsevencoll2}
\cR_7^\mathrm{WL}\left(u_2,u_3,\sqrt{u_1/\kappa},
\frac{1-\sqrt{u_1/\kappa}}{1-u_1},0,
\frac{1-\sqrt{u_1\kappa}}{1-u_1},\sqrt{u_1\kappa}\right) = \cR_6^\mathrm{WL}
(u_1,u_2,u_3)
\ ,
\eeq
and note that the left hand side must therefore be independent of the variable
$\kappa$. This can be thought of as the $z$-independence of $\cR_7^\mathrm{WL}$ in the collinear limit -- which is precisely the feature
one would expect from the scattering amplitude.

\begin{figure}[ht]
\begin{center}
\includegraphics[width=12cm]{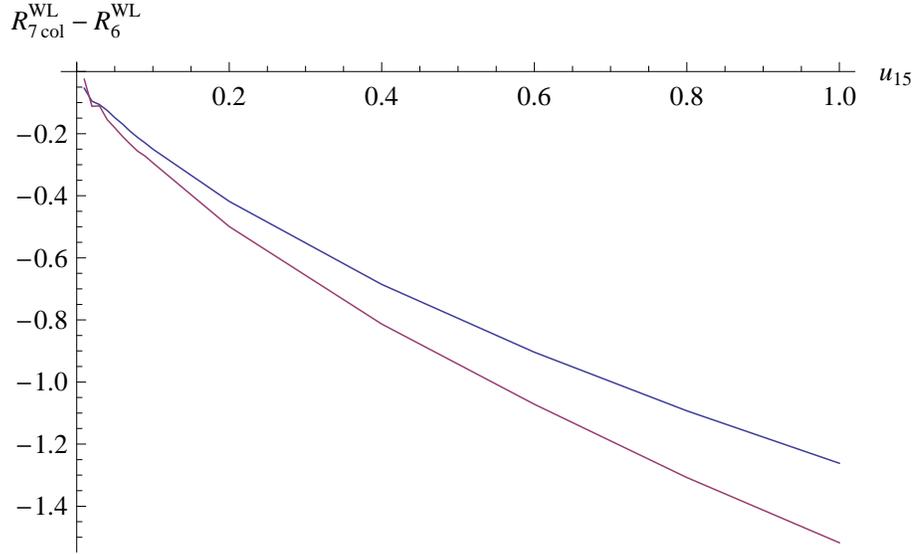}
\end{center}
\caption{\it A plot of the seven-point remainder function minus the
  six-point remainder function in the collinear limit. The kinematics are
  given in~\eqref{Rsevencoll2} with $(u_1,u_2,u_3)=(1/10,1,1)$ but
  with a non-vanishing $u_{15}$ which varies along the x-axis
  ($u_{15}$ vanishes in the collinear limit). For the blue graph we
  have $\kappa=0.25$ and for the purple graph $\kappa=0.49$. We see
  that in both cases
  the difference tends to zero  with $u_{15}$,  confirming~\eqref{Rsevencoll2}.    }
\label{fig:7pntcol}
\end{figure}

We have computed $\cR_7^\mathrm{WL}$ in the collinear limit and have confirmed that \eqref{Rsevencoll2} does hold
within the errors and that no dependence on $\kappa$ is found (see
Table~\ref{table7pointcol}
and Figure~\ref{fig:7pntcol} for example). Again, we stress the
absence on the right hand side of
\eqref{Rsevencoll2} of any additional constant term.

\begin{table}[h]
\begin{center}
\begin{tabular}{|c||c|c|c|c}
\hline
$(u_1,u_2,u_3)$ &
$\cR^\mathrm{WL}_{7 \mathrm{col}}(\kappa=2.5)$ &
$\cR^\mathrm{WL}_{7 \mathrm{col}}(\kappa=4.9)$ &
$\cR^\mathrm{WL}_6$  \\
\hline
\hline
$(1/10,1,1)$   & -2.78972&-2.76053&-2.73441 \\
\hline
\end{tabular}
\end{center}
\caption{\it Checks of the collinear limit $\cR^\mathrm{WL}_7
    \rightarrow \cR^\mathrm{WL}_6$. We present
$\cR^\mathrm{WL}_{7\rm{col}}(\kappa):=\cR_7^\mathrm{WL}\big(u_2,u_3,
  \sqrt{u_1/\kappa},
\frac{1-\sqrt{u_1/\kappa}}{1-u_1},0.01,
\frac{1-\sqrt{u_1\kappa}}{1-u_1},\sqrt{u_1\kappa}\big)$  for
different values of $\kappa$, together with its collinear limit
$\cR_6(u_1,u_2,u_3)$ for $(u_1,u_2,u_3)=(1/10,1,1)$. Within numerical errors
 the values  agree.}
\label{table7pointcol}
\end{table}

In the following section we compare these results with what can be learned from the multi-collinear limits.

\subsection{Multi-collinear limits}

Here we would like  to derive the multi-collinear equivalent of the general reduction formulae in
\eqref{abb}, \eqref{Rsevencoll}-\eqref{Rsevencoll2}
for $n$-gon Wilson loops.

The first non-trivial case is a triple collinear limit of a six-point configuration considered in Section 5 of \cite{seven}.
In the limit where $p_4$, $p_5$ and $p_6$ become  collinear one has
\beq
\label{tripcol567sixpoints}
p_4 := x_4-x_5 = z_1P \  , \quad p_5 = x_5-x_6 = z_2 P \  , \quad p_6 = x_6-x_1 = z_3 P \ , \quad
z_1+z_2+z_3 =1 \ .
\eeq
In this limit, the two-loop scalar function $\cM^{(2)}_6$ behaves as \cite{seven}
\beq
\label{M23r}
\cM^{(2)}_6 \to \cM_4^{(2)} \, + \, \cM_4^{(1)}\, r_S^{(1)}  \Big( {s_{45} \over s_{456}} ,  {s_{56} \over s_{456}}, z_1 , z_3, \e \Big) \, +\,
 r_S^{(2)}  \Big( {s_{45} \over s_{456}} ,  {s_{56} \over s_{456}}, z_1 , z_3, \e \Big)
\ ,
\eeq
where $ r_S^{(1)} $ and $r_S^{(2)} $ are the one- and two-loop triple  splitting amplitudes.
The two-loop triple splitting amplitude does not satisfy an iteration relation similar to the
that found in \cite{abdk} for the simple splitting amplitude \eqref{babis2} \cite{seven}.
It can however be decomposed into
a term with does, plus a two-loop finite remainder, as
\beq
\label{r3bdsr}
 r_S^{(2)}
\ := \
 r_S^{(2)\, \mathrm{BDS} }   \, + \,
 \widetilde{\cR}
 \ .
 \eeq
It was shown in \cite{seven} that $\widetilde{\cR}$ is nothing but the finite remainder function at six points
 evaluated in the triple collinear kinematics.
 Indeed, in the triple collinear limit  the six-point cross-ratios
remain independent and do not vanish, $u_i \to \bar{u}_i$, $i=1,2,3$,
where
\beq
\label{ubars}
\bar{u}_1 =  {1 \over 1-z_3}{s_{45}  \over s_{456} } \ , \quad
\bar{u}_2 =  {1 \over 1-z_1}{s_{56}  \over s_{456} } \ , \quad
\bar{u}_3 = {z_1z_3 \over (1-z_1)(1-z_3)} \ .
\eeq
Hence,
\beq
\label{sss}
\cR_6(u_1, u_2, u_3)  \to \cR_6(\bar{u}_1, \bar{u}_2, \bar{u}_3)
\ .
\eeq
If we now take
the triple collinear limit of the remainder function as defined  from \eqref{fm2bis}, and use
\eqref{M23r} and \eqref{r3bdsr}, we get
 \beq
 \label{ccc}
 \cR_ 6  (u_1 , u_2 , u_3 ) \ \to \ \cR_4 \, + \,  \widetilde{\cR} (\bar{u}_1, \bar{u}_2, \bar{u}_3)
 \ .
 \eeq
Using that  $ \cR_4  = 0$, and comparing \eqref{ccc} to \eqref{sss}  we get at once that
\beq
\label{ddd}
 \widetilde{\cR} (\bar{u}_1, \bar{u}_2, \bar{u}_3) \ = \ \cR_6 (\bar{u}_1, \bar{u}_2, \bar{u}_3)
 \ .
 \eeq
Assuming further dual conformal invariance, the conclusion of \cite{seven} is
that determining the remainder function  $\cR_6$ in the triple collinear limit
is equivalent to determining it in full generality.

We can now derive the triple collinear limit
of the two-loop $n$-point remainder function for general $n\ge 6$. Using \eqref{ddd}
we have
 \beq
 \label{ccc2}
 \cR_ n  \ \to \ \cR_{n-2} \, + \,  \cR _6(\bar{u}_1, \bar{u}_2, \bar{u}_3)
 \ ,
 \eeq
where $(\bar{u}_1, \bar{u}_2, \bar{u}_3)$ are still defined by \eqref{ubars} (in the case where
the collinear momenta are
$p_4$, $p_5$ and $p_6$).

What about Wilson loops? Similarly to our discussion of simple collinear limits,
we expect that, for Wilson loops, the triple collinear limit of the Wilson loop remainder function is given by
 \beq
 \label{cccw2}
 \cR_ n^\mathrm{WL}\left( u_{14},u_{25},u_{36},u_{47},u_{15},u_{26},u_{37}\right)
\ \rightarrow \ \cR_{n-2}^\mathrm{WL} \, + \,  \cR_6^\mathrm{WL} (\bar{u}_1, \bar{u}_2, \bar{u}_3)
 \ .
 \eeq
Now consider a triple collinear limit of a  heptagon Wilson loop, where $p_5$, $p_6$ and $p_7$ are collinear,
\beq
\label{tripcol567}
p_5 = x_5-x_6 = z_1P \  , \quad p_6 = x_6-x_7 = z_2 P \  , \quad p_7 = x_7-x_1 = z_3 P \ , \quad
z_1+z_2+z_3 =1 \ .
\eeq
In this limit,
\beq
 \cR_ 7^\mathrm{WL}  \ \to \ \cR_{6}^\mathrm{WL} (\bar{u}_1, \bar{u}_2, \bar{u}_3)
 \ ,
 \eeq
since there is no five-point remainder function.

For our present case of collinear $p_5$, $p_6$ and $p_7$ the variables
$\bar{u}_i$ read
\beq
\label{ubarsaltro}
\bar{u}_1 =  {1 \over 1-z_3}{x_{57}^2 \over x_{15}^2} \ , \quad
\bar{u}_2 =  {1 \over 1-z_1}{x_{16}^2 \over x_{15}^2} \ , \quad
\bar{u}_3 = {z_1z_3 \over (1-z_1)(1-z_3)} \ .
\eeq
The triple collinear limit \eqref{tripcol567} of the seven-point kinematics gives
\bea
&&x^2_{15} \sim x^2_{16} \sim x^2_{57}  \to 0 \ , \nonumber \\
&&x^2_{27} = z_3 x^2_{25} \ , \quad  x^2_{26} = (1-z_1) x^2_{25} \ , \quad
x^2_{46} = z_1 x^2_{41} \ , \quad  x^2_{47} = (1-z_3) x^2_{41} \ , \nonumber \\
&& x^2_{36} = (1-z_1)x^2_{35} +  z_1 x^2_{13} \ , \quad x^2_{37} = (1-z_3)x^2_{13} +  z_3 x^2_{35} \ ,
\eea
with all other $x^2_{ij}$ segments unmodified.
The  cross-ratios $u_{ij}$ take the form:
\bea
\label{useven3col}
&&u_{14} =  0 \ , \quad
u_{25} = {(1-z_1)x_{35}^2 \over z_1 x_{13}^2 + (1-z_1)x_{35}^2}  \nonumber
 \\
&&u_{36} = {z_1 \over 1-z_3}{z_3 x_{35}^2 + (1-z_3)x_{13}^2 \over z_1 x_{13}^2 + (1-z_1)x_{35}^2}  \ , \quad
u_{47} =  {1 \over 1-z_3}{x_{57}^2 \over x_{15}^2} \ , \quad
u_{15} =   {1 \over 1-z_1}{x_{16}^2 \over x_{15}^2} \ ,  \nonumber \\
&&u_{26}=  {z_3 \over 1-z_1}{z_1 x_{13}^2 + (1-z_1)x_{35}^2 \over z_3 x_{35}^2 + (1-z_3)x_{13}^2} \ ,
\quad u_{37} =  {(1-z_3)x_{13}^2 \over z_3 x_{35}^2 + (1-z_3)x_{13}^2}
\ .
\eea
These relations imply two things. First we note that there are three constraints on the seven variables,
\beq
\label{constrs3}
u_{14} =  0 \ , \quad
u_{25} + u_{36} u_{37} = 1 \ , \quad  u_{37} + u_{25} u_{26} =1 \ ,
\eeq
leaving four variables unconstrained, out of which the three conformal ratios coincide with the $\bar{u}_i$
variables of  \eqref{ubarsaltro},
\beq
\label{constrs4}
u_{47} =  \bar{u}_1 \ , \quad
u_{15} =\bar{u}_2 \ , \quad  u_{26}u_{36}=\bar{u}_3 \ .
\eeq
Taking this into account, and solving the two constraints in \eqref{constrs3} for $u_{14}$ and $u_{35}$ we
conclude that the triple collinear limit relates the seven-point remainder function to the six-point one
as follows:
\beq
\label{Rseven3coll}
\cR_7^\mathrm{WL}\left(0,
\frac{1-u_{36}}{1-u_{26}u_{36}}, u_{36},u_{47},u_{15},u_{26},\frac{1-u_{26}}{1-u_{36}u_{26}}\right) =\, \cR_6^\mathrm{WL} (u_{47},u_{15},u_{26}u_{36})
\ .
\eeq
Notice that since $\cR_7^\mathrm{WL}$ is invariant under cyclic interchange of its variables, this is exactly the same equation as we had before in the simple collinear limit \eqref{Rsevencoll} (alas expressed in terms of
$u_{36}$, $u_{47}$, $u_{15}$ and $u_{26}$).

We conclude that  the simple and the triple collinear limits described by
\eqref{Rsevencoll} and \eqref{Rseven3coll} give identical information about $\cR_7^\mathrm{WL}$.
As before, the left hand side of \eqref{Rseven3coll}
cannot depend on one particular combination of cross-ratios, namely  $u_{36}/u_{26} :=\kappa$.
Thus, we can rewrite \eqref{Rseven3coll} as
\beq
\cR_7^\mathrm{WL}\left(0,
\frac{1-\sqrt{\bar u_{3}\kappa}}{1-\bar u_{3}},\sqrt{\bar u_{3} \kappa},\bar u_{1}, \bar u_{2},\sqrt{\bar
    u_{3}/\kappa},
\frac{1-\sqrt{\bar u_{3}/\kappa}}{1-\bar u_{3}}\right) = \, \cR_6^\mathrm{WL} (\bar
u_{1},\bar u_{2},\bar u_{3})
\ ,
\eeq
and note that the left hand side must be independent of the variable
$\kappa$.

Finally, we have investigated the quadruple collinear limit, which is the highest non-trivial multi-collinear
limit one can take on the seven-point kinematics. We have found in this limit that all seven cross-ratios
are (a) mutually independent, and (b) are expressed entirely in terms of the multi-collinear kinematics
(i.e.~they are functions of $z_1, \ldots, z_4$ and ratios of kinematic invariants involving only the collinear
momenta). In this way, the quadruple collinear limit does not add any non-trivial functional constraints on the
$\cR_7$, however, it elucidates its physical (scattering amplitudes-based) meaning,
\beq
\cR_7^\mathrm{WL}( u_{14},u_{25},u_{36},u_{47},u_{15},u_{26},u_{37}) \,
\, = \,
\Delta {\rm split}_4 :=\, r_{S4}^{(2)} - r_{S4}^{(2)\, {\rm BDS}}\ ,
\eeq
where $\Delta {\rm split}_4$ is the normalised two-loop level part of the quadruple splitting function $r_{S4}^{(2)}$
which is not already accounted by the BDS contribution $r_{S4}^{(2)\, {\rm BDS}}$.
Hence, similarly to the six-point case discussed in previously \cite{seven}, we see that the remainder function $\cR_7$
is entirely determined by the quadruple splitting function.

\section{Eight-point Wilson loops and beyond}

It is natural to seek beyond the encouraging results at seven points given above,
and see if these persist at eight points. The discussion earlier indicates that
there are twelve independent conformal cross-ratios in this case
(we take the external momenta to be on shell and do not impose the Gram determinant constraint),
whereas there are twenty independent momentum invariants. These
independent invariants may be taken to be
\beq
\label{20invariants}
x^2_{i\; i+2},x^2_{i+4 \; i+6}, \; x^2_{i \;i+3}, \; x^2_{i+4\; i+7},\; x^2_{i\;i+4} \ ,
\quad{i=1,\ldots,4}
\ .
\eeq
We will use the following twelve cross-ratios:
\beq
\label{12crossratios}
u_{i\; i+3} \ , \quad i=1,\ldots,8\ ,
\qquad
u_{i\;i+4} \ , \quad i=1, \ldots,4 \ ,
\eeq
and label these $u_1,...,u_{12}$. Instead of the twenty momentum invariants given in \eqref{20invariants}, we will use these twelve cross-ratios, plus the following eight momentum invariants
\beq
\label{other8}
x^2_{i+5 \; i+8},\; x^2_{i\;i+4} \ , \quad i=1,\dots,4\ ,
\eeq
which we will call $m_1,\dots,m_8$.
The remaining momentum invariants $x^2_{i\; i+2}, x^2_{i+4\; i+6}$ and  $x^2_{i+1\; i+4}$, for $i=1,\dots,4$, are then dependent variables.

The first question to study is whether the eight-point Wilson loop remainder function $\cR_8^\mathrm{WL}$ is
only a function of the twelve conformal cross-ratios \eqref{12crossratios}, and not
of the additional eight invariants \eqref{other8}. To do this, one may fix
a choice of the cross-ratios, then calculate the eight-point remainder function
for various choices of kinematics, corresponding to different choices of the
variables \eqref{other8}. For example, in Table  \ref{tab:8point1}
we have listed some numerical results for the case of all the cross-ratios of \eqref{12crossratios} equal to one.
We find similar results for more generic values of the
cross-ratios -- an example is given in Table \ref{tab:8point1bis}.

\begin{table}[th]
\begin{center}
\begin{tabular}{|c|c|}
\hline $(m_1,\dots,m_8)$ &
 $\cR_8^\mathrm{WL}$
\\
\hline  $(-1,-1,-1,-1,-1,-1,-1,-1)$ & -4.603
\\
\hline   $(-2,-2,-2,-2,-2,-2,-2,-2)$ & -4.602
\\
\hline $(-1,-2,-4,-8,-1,-2,-4,-8)$ & -4.605
\\
\hline $(-5,-3,-5,-3,-1,-3,-5,-7)$  & -4.605
\\
\hline
\end{tabular}
\end{center}
\caption{\it The remainder function $\cR_8^\mathrm{WL}$ for $u_1=u_2=\dots =u_{12}=1$ and different choices of the other independent invariants \eqref{other8}. The errors in $\cR_8^\mathrm{WL}$ are approximately $0.02$.}
\label{tab:8point1}
\end{table}
\begin{table}[th]
\begin{center}
\begin{tabular}{|c|c|}
\hline $(m_1,\dots,m_8)$ &
 $\cR_8^\mathrm{WL}$
\\
\hline   $(-2,-3,-4,-1,-5,-6,-7,-8)$ & 5.993
\\
\hline  $(-1/3,-1/4,-1/9,-1/2,-1/8,-1/7,-1/6,-1)$ & 5.984
\\
\hline
\end{tabular}
\end{center}
\caption{\it The remainder function $\cR_8^\mathrm{WL}$ for the choice of cross-ratios $(u_1,\dots,u_{12})
= (2,3,4,1/2,1/3,1/4,1/5,1,1/5,1/6,1/7,1/8)$ and two choices of the other independent invariants \eqref{other8}. The errors in $\cR_8^\mathrm{WL}$ are approximately $0.04$.}
\label{tab:8point1bis}
\end{table}

These results support the conjecture that the eight-point remainder
function $\cR_8^\mathrm{WL}$ is a function of the twelve cross-ratios \eqref{12crossratios}
alone, and not of any other independent additional momentum invariants.

Given this, a second question concerns symmetries of the function $\cR_8^\mathrm{WL}(u_1,...,u_{12})$. The Wilson loop is invariant under cyclic permutations of the
external momenta as well as under parity. For the case of eight points, this
implies that the remainder function $\cR_8^\mathrm{WL}(u_1,...,u_{12})$ should be
invariant under cyclic permutations of the first eight and last four cross-ratios simultaneously, as well as being invariant under the simultaneous reversal of both the
first eight and last four cross-ratios. We find numerical agreement with this -- for example,
\bea
\label{cyclicsymmetry1}
&& \cR_8^\mathrm{WL}(2,2,1,1,1,1,1,1,1,1,1,1)= -3.712, \nonumber
 \\
&& \cR_8^\mathrm{WL}(1,2,2,1,1,1,1,1,1,1,1,1)= -3.712,
\eea
with errors $\sim 0.02$. We have also seen numerically in a number of cases that there is no invariance under more general permutations of the cross-ratios.

Let us now consider collinear limits of the eight-point remainder function.
As an example we take $p_7$ and $p_8$ to be collinear. We set
\beq
\label{coll78}
p_7 = x_7-x_8 = z P \  , \qquad p_8 = x_8-x_1 = (1-z) P \ ,
\eeq
where as usual $P^2 \to 0$ in the collinear limit. In this collinear limit, the cusp at $x_8$ is  ``flattened".
In this limit we find that the eight-point cross-ratios $(u_{14},u_{25},u_{36},u_{15},u_{26})$ reduce directly to the cross-ratios in the
seven-point case with the same names, $u_{16}\rightarrow 0$ and
the seven-point cross-ratios $(u_{47}, u_{37})
$ are given in terms of the
eight-point ones as $(u_{47}u_{48},u_{37}u_{38})$ respectively.
Finally, we have the following relations amongst the eight-point cross-ratios:
\bea
\label{8pointrelations}
&& u_{27}u_{37}u_{38} = -1 + u_{27} + u_{38}, \nonumber
 \\ [4pt]
&& (-1 +u_{27}+u_{38})u_{47}= u_{38}(1- u_{58}), \nonumber \\ [4pt]
&& u_{48}u_{58} = 1- u_{27}u_{37},\nonumber \\ [4pt]
&& u_{37}(-u_{38}-u_{47}+u_{38}u_{47}+u_{38}u_{58}) = -1 + u_{58},
\eea
which are solved by
\bea
\label{8pointsolutions}
&& u_{27} = \frac{-1  + u_{38}}{-1 + u_{37}u_{38}}, \nonumber
  \\ [8pt]
&& u_{58} = \frac{ -1 + u_{37}u_{38} + u_{37}u_{47} - u_{37}u_{38}u_{47}} { -1 + u_{37}u_{38} }, \nonumber \\ [8pt]
&&  u_{48} =   \frac { -1+ u_{37}  }   {  -1 + u_{37}u_{38} + u_{37}u_{47} - u_{37}u_{38}u_{47}  } .
\eea
The three variables $(u_{37}, u_{38}, u_{47})$ in the above are then freely specifiable.

Analysis of the remainder function given earlier
implies that in the collinear limit
\beq
\label{sshift88}
 \cR_8^{\mathrm{WL}} \rightarrow   \cR_7^{\mathrm{WL}}\ .
\eeq
Hence in the collinear limit \eqref{coll78} one should have
\bea
 \label{Reightcoll}
&&\cR_8^\mathrm{WL}( u_{14},u_{25}, u_{36},u_{47},u_{58},u_{16},
u_{27}  ,u_{38},u_{15},u_{26},u_{37},u_{48}) \rightarrow
\nonumber \\ [6pt]
&&  \cR_8^\mathrm{WL}\left(u_{14},u_{25}, u_{36},u_{47},u_{58}^*,0,
u_{27}^*  ,u_{38},u_{15},u_{26},u_{37},u_{48}^*\right) \nonumber\\ [6pt]
&&= \cR_7^\mathrm{WL}  (u_{14},u_{25},
u_{36},u_{47}u_{48}^*,u_{15},u_{26}, u_{37}u_{38}) \ ,
\eea
where the stars in the above indicate that the solutions
\eqref{8pointsolutions} are to be
inserted.

One can test this directly. For example, for the choices
of values for the independently specifiable eight-point cross-ratio
variables
\beq
(u_{14},u_{25}, u_{36},u_{47},u_{83},u_{15},u_{26},u_{37})
= (1,1,1,1,1/2,1,1,1/2),
\eeq
 and taking $u_{61} = 0.001$ one finds
\bea
 \label{Reightcolltest}
&&  \cR_8^\mathrm{WL}(u_{14},u_{25}, u_{36},u_{47},u_{58}^*,0.001,
u_{72}^*  ,u_{83},u_{15},u_{26},u_{37},u_{48}^*)  = -4.2756, \nonumber\\
&&\cR_7^\mathrm{WL}  (u_{14},u_{25}, u_{36},u_{47}u_{48},u_{15},u_{26},
u_{37}u_{38}) = -4.2906\ ,
\eea
with errors of $0.147$ and $0.005$ respectively.

The success of the above numerical tests of the conformal symmetry,
functional symmetries and collinear limits of the eight-point remainder
function supports the conjecture
that the Wilson loop is correctly reproducing the physical
amplitude at this level.

There are no further conceptual or computational obstacles to generalising the
above work beyond eight-point Wilson loops, apart from the question of the computer time required
to numerically calculate the integrals -- we stress that no new
integrals arise in the Wilson loop approach to $n$-point two-loop diagrams for any $n$, apart from those which we have
already discussed, and we are able to calculate all the diagrams introduced for
generic values of the momentum variables $Q_i$ (numerically, and in a number of cases, analytically).

This means that we have full numerical control over two-loop $n$-gon Wilson loops and,
if the correspondence with amplitudes continues to hold, over $n$-point MHV amplitudes at arbitrary $n$ in the planar ${\cal N}=4$ theory. This should be contrasted with the situation where one calculates amplitudes directly.

\vspace{1cm}
\section*{Acknowledgements}

It is a pleasure to thank Mert Aybat, Zvi Bern, Lance Dixon, James Drummond, George Georgiou,
Nigel Glover, David Kosower, Lorenzo Magnea  and
Marcus Spradlin for discussions,
and especially Lance Dixon for useful comments and Radu Roiban for discussions and initial collaboration on this project.
PH, VVK and GT would also like to thank the ETH Z\"{u}rich for hospitality and support
during the earlier stage of this project.
This work was supported by the Swiss National Science Foundation under
contract 200021-117873, and by the STFC under the
Queen Mary Rolling Grant ST/G000565/1 and the IPPP Grant ST/G000905/1.
The work of PH is supported by an EPSRC Standard Research Grant EP/C544250/1.
VVK acknowledges a Leverhulme Research Fellowship and
GT is supported by an EPSRC Advanced Research Fellowship EP/C544242/1
and by an EPSRC Standard Research Grant EP/C544250/1.

\newpage

\startappendix

\section{A note on conventions}

Wilson loops are computed in dimensional reduction  in $D=4-2\epsuv$ dimensions with $\epsuv >0$
to regularise the UV divergences. To facilitate the comparison with scattering amplitudes (which require infrared regularisation)
we introduce
\beq
\label{epsUV}
\eps = - \epsuv \ .
\eeq
The perturbative expansion of the Wilson loop is characterised by \eqref{nae}, \eqref{w2our}:
\bea
\label{naeA}
\lan W[\cC_n ] \ran   &=&  1 \, + \, \sum_{l=1}^{\infty} a^l W^{(l)}_n \ = \   \exp \sum_{l=1}^{\infty} a^l w^{(l)}_n \ , \\
\label{w2ourA}
w^{(2)}_n &=&  W^{(2)}_n \, - \, {1\over2} \, (W^{(1)}_n)^2
\ ,
\eea
and \eqref{no2} defines the Laurent expansion in $\eps$ for the two-loop contribution,
\beq
\label{no2A}
w^{(2)}_n \ = \
\sum_{i=1}^n { \left( -{ x_{i i+2}^2\over \m^2}\right)^{-2\eps} } \Big(
{ w^{(2)}_{-2} \over \eps^2} +
{w^{(2)}_{-1} \over \eps}  \Big)  \ + \ F^{(2)}_{n} \, + \, \cO(\e )
\ ,
\eeq
where $F^{(2)}_{n}$ is the finite part of the Wilson loop.

In \cite{dhks6}  the exponent of the Wilson loop in \eqref{naeA} was defined as
$a w^{(1)} + 2 a^2 w^{(2)} + \cdots$,  and thus a corresponding factor of $1/2$
would need to be introduced in front of the right hand side in \eqref{w2ourA}
and in \eqref{no2A} if we were to switch to their conventions.
Therefore our singular terms
$w_{-2}$, $w_{-1}$  and the finite part $F^{(2)}_{n}$ in \eqref{no2A} are related to the
$A_{-2}$, $A_{-1}$ and $A_{0}$ contributions of \cite{dhks} as follows:
\beq
\label{wvsAs}
w_{-2} = \sum_\alpha A_{-2}^{(\alpha)} \ , \qquad
w_{-1} = -\sum_\alpha A_{-1}^{(\alpha)} \ , \qquad
F^{(2)}_{n} = 2 \sum_\alpha A_{0}^{(\alpha)} \ .
\eeq
The minus sign in the second equation is due to \eqref{epsUV}.

\section{Most general hard diagram}

Below is the result for the diagram where a three-point vertex is attached to three
lightlike momenta $p_1$, $p_2$, $p_3$ of the Wilson loop, which we call the ``hard diagram" as it is the most difficult to evaluate analytically in general.
These momenta are separated by
the three, not necessarily lightlike momenta, $Q_3$, $Q_1$, $Q_2$,
where $Q_3$ is between $p_1$ and $p_2$ and so on (see Figure \ref{harddiagram}).
Momentum conservation is then $\sum_{i=1}^{3} (p_i + Q_i ) = 0$.
We also set $D = 4 - 2 \epsuv= 4 + 2 \eps$ where $\epsuv = - \eps >0$.
The special four-point case is considered later.

We write this diagram in the most general configuration as%
\footnote{We remind the reader that we will always
suppress the common prefactor defined in \eqref{multby}
from the expression of all diagrams.
}
\begin{align}
&  f_H(p_1,p_2,p_3;Q_1,Q_2,Q_3)\nonumber \\
&:= {1 \over 8}\,{\Gamma (2-2\epsuv) \over \Gamma(1-\epsuv)^2}
\int_{0}^{1}\! \Big(  \prod_{i=1}^{3} d\t_i \Big) \int_{0}^{1}\!\Big(\prod_{i=1}^{3} \, d\alpha_i \Big)\delta ( 1 - \sum_{i=1}^3
\alpha_i )  \ (\alpha_1 \alpha_2 \alpha_3)^{-\epsuv}  { \mathcal{N} \over \mathcal{D}^{2-2\epsuv}}\ ,
\end{align}
where
\beq \label{D}
\mathcal{D} := -\a_1 \a_2 (z_1 - z_2)^2 -\a_2 \a_3 (z_2 - z_3)^2 -\a_1 \a_3 (z_1 - z_3)^2 \ ,
\eeq
and
\beqa
(z_1 - z_2)^2 &=& Q_3^2 + 2 (p_1 p_2) (1 - \t_1) \t_2 + 2 (Q_3 p_1 ) ( 1 - \t_1) + 2 (Q_3 p_2 ) \t_2
\ ,
\\ \nonumber
(z_2 - z_3)^2 &=& Q_1^2 + 2 (p_2 p_3) (1 - \t_2) \t_3 + 2 (Q_1 p_2 ) ( 1 - \t_2) + 2 (Q_1 p_3 ) \t_3
\ ,
\\ \nonumber
(z_3 - z_1)^2 &=& Q_2^2 + 2 (p_3 p_1) (1 - \t_3) \t_1 + 2 (Q_2 p_3 ) ( 1 - \t_3) + 2 (Q_2 p_1 ) \t_1
\ .
\eeqa
The original expressions for the $z_i - z_{i + 1}$ are
\beqa
z_i - z_{i+1}  & = & Q_{i+2} + p_i (1 - \t_i) + p_{i+1} \t_{i+1} \ , \qquad  i=1, 2,3
\  .
\eeqa
The expression for the numerator $\cN$ has two kinds of terms. The first three lines involve
$\t$ and $\a$ parameters, whereas the remaining three lines involve only the $\t$ parameters.
It is given by
\beqa
\label{N}
\cN &=&  2 (p_1 p_2) (p_1 p_3) \Big[ \a_1 \a_2 ( 1 - \t_1) +  \a_3 \a_1 \t_1 \Big]
\cr
&+&  2 (p_1 p_2) (p_2 p_3) \Big[ \a_2 \a_3 ( 1 - \t_2) + \a_1 \a_2 \t_2 \Big]
\cr
&+&  2 (p_1 p_3) (p_2 p_3) \Big[ \a_3 \a_1 ( 1 - \t_3) + \a_2 \a_3 \t_3 \Big]
\cr
&+& 2 \a_1 \a_2 \Big[ 2 (p_1 p_2) (p_3 Q_3) - (p_2 p_3) (p_1 Q_3) - (p_3 p_1 ) (p_2 Q_3) \Big]
\cr
&+& 2 \a_2 \a_3 \Big[ 2 (p_2 p_3) (p_1 Q_1) - (p_3 p_1) (p_2 Q_1) - (p_1 p_2 ) (p_3 Q_1) \Big]
\cr
&+& 2 \a_3 \a_1 \Big[ 2 (p_3 p_1) (p_2 Q_2) - (p_1 p_2) (p_3 Q_2) - (p_2 p_3 ) (p_1 Q_2) \Big]
\ .
\eeqa

\subsection{Four-point case}
The four-point case  can be obtained by setting
\beq
Q_3 = Q_1 = 0 \ , \qquad Q_2 = p_4 = - (p_1 + p_2 + p_3) \ ,
\eeq
where now $Q_2^2 = p_4^2 = 0 $. The expression for $\cN$ in \eqref{N} then simplifies to
\beqa
\cN & = & 2 (p_1 p_2) (p_1 p_3) \,  ( 1 - \t_1)   \a_1 (\a_2 - \a_3)
\cr
&+&  2 (p_1 p_2) (p_2 p_3) \Big[ \a_2 \a_3 ( 1 - \t_2) + \a_1 ( 2 \a_3 + \a_2 \t_2 )  \Big]
\cr
&+&  2 (p_1 p_3) (p_2 p_3) \, \t_3 \a_3 ( \a_2 - \a_1)
\ .
\eeqa
In this special case we have
\beqa
(z_1 - z_2)^2 &=&  2 (p_1 p_2) (1 - \t_1) \t_2
\ ,
\\ \nonumber
(z_2 - z_3)^2 &=&  2 (p_2 p_3) (1 - \t_2) \t_3
\ ,
\\ \nonumber
(z_3 - z_1)^2 &=&  2 (p_3 p_1) (1 - \t_3) \t_1 + 2 (p_3 p_4 ) ( 1 - \t_3) + 2 (p_1 p_4 ) \t_1
\ ,
\eeqa
where we can set
\beqa
s&:=& 2 (p_1 p_2 ) = 2 (p_3 p_4) \ , \cr
t&:=& 2 (p_2 p_3 ) = 2 (p_1 p_4) \ , \cr
u&:=& 2 (p_1 p_3 ) = 2 (p_2 p_4) \ ,
\eeqa
and $s +  t + u = 0$.

The denominator in the four-point case then simplifies to
\beq
\cD :=s \,  \a_1 \a_2  \t_2 ( 1 - \t_1) + t\, \a_2 \a_3 \t_3 ( 1 - \t_2) + \a_3 \a_1 \Big[ s ( 1 - \t_1) ( 1 - \t_3) + t \t_1 \t_3 \Big]
\ .
\eeq

Notice that in the six-point case there is a new diagram where all the $Q_i$'s are made of a single
lightlike momentum; in this diagram  $Q_1^2 = Q_2^2 = Q_3^2 = 0$.
One should find that the result for the corresponding integral $\cI$ is {\it finite} in four dimensions.

\section{Curtain diagram}
We call ``curtain" diagrams those diagrams where two propagators connect three different edges, as depicted in
Figure \ref{curtaindiagram}.
We use the same notation as in the three-point vertex case above, with propagators stretching from
$p_1$ to $p_2$ (with end-points $z_1(\sigma_1)$ and $z_2(\tau_2)$)  and from $p_1$ to $p_3$  (with end-points $z_1( \t_1)$ and $z_3(\t_3)$).

Just from looking at the diagram we should have the following symmetry: $p_2 \leftrightarrow p_3$ $Q_2 \leftrightarrow Q_3$.
We have
\begin{align}
z_1(\t_1)&=p_1\t_1\\
z_1(\s_1)&=p_1\s_1\\
z_2(\t_2)&=p_2\t_2+p_1 +Q_3\\
z_3(\t_3)&=-p_3\t_3-Q_2\ .
\end{align}
The exponentiation theorem says we should only consider the diagram where the internal gluon propagators cross -  this gives the constraint $\t_1 >\s_1$.

The diagram represents the following contribution to the Wilson loop:
\beq
-{1\over 2}   \int_0^1\!\Big(  \prod_{i=1}^{3} d\t_i \Big)\,\int_0^{\t_1}d\s_1 {(p_1p_2)
\over \big( -[z_1(\s_1)-z_2(\t_2)]^2 \big)^{1-\epsuv}}\,{(p_1p_3) \over
\big( -[z_1(\t_1)-z_3(\t_3)]^2 \big)^{1-\epsuv}}\ .
\eeq
Putting in the values for the end-points we get the following integral representation:
\begin{align}
-{1\over 2} \,\int_0^1\! \Big(  \prod_{i=1}^{3} d\t_i \Big)\,\int_0^{\t_1}d\s_1 &{(p_1p_2) \over \big[ -2(p_1Q_3)
(1-\s_1)-2(p_1p_2)(1-\s_1)\t_2-2(p_2Q_3)\t_2-Q_3^2 \big]^{1-\epsuv}} \nonumber \\
&\times {(p_1 p_3) \over
\big[  -2(p_1Q_2) \t_1-2(p_1p_3)\t_1\t_3-2(p_3Q_2)\t_3-Q_2^2 \big]^{1-\epsuv}} \ .
\end{align}
A more symmetrical way to write this is to send $\s_1 \rightarrow
1-\sigma_1$ in which case the constraint $\t_1 >\s_1$ becomes $\t_1
+\s_1 >1$ and the integrand would be manifestly symmetric under $p_2
\leftrightarrow p_3$, $Q_2 \leftrightarrow Q_3$, $\t_1 \leftrightarrow
\s_1$ and $\tau_2 \leftrightarrow \tau_3$. We have performed this
change of variables to obtain~(\ref{curtain}).

\section{Cross diagram (involving two sides)}

This diagram consists of two gluon propagators, stretching from sides $p_1$ to $p_2$ with sides $Q_1$ and $Q_2$ between,
as represented in  Figure \ref{crossdiagram}.

The end-points of the first propagator are $z_1(\tau_1)$ and $z_2(\tau_2)$, and of  the second are $z_1(\sigma_1)$ and $z_2(\sigma_2)$, with
\begin{align}
  z_1(\sigma_1)&=p_1 \sigma_1 \qquad  z_2(\sigma_2)=-p_2 \sigma_2 -
  Q_2  \nonumber \\
  z_1(\tau_1)&=p_1 \tau_1 \qquad  z_2(\tau_2)=-p_2 \tau_2 - Q_2 \ .
\end{align}
In order to ensure the crossing of the propagators we require $\tau_1 <\sigma_1$
and $\tau_2 >\sigma_2$.
The diagram then represents the  integral
\begin{align}
-{1\over 2} \, \int_0^1 d \sigma_1 d\tau_2
 \int_0^{\sigma_1}d\tau_1 \int_0^{\tau_2} d\sigma_2 {(p_1p_2) \over
   \big(-[z_1(\sigma_1)-z_2(\sigma_2)]^2\big)^{1-\epsuv}}
 {(p_1p_2) \over
   \big(-[z_1(\tau_1)-z_2(\tau_2)]^2\big)^{1-\epsuv}}\ .
\end{align}
Putting in the values of the end-points gives the integral
\begin{align}
-{1\over 2} \, \int_0^1 d \sigma_1 d\tau_2
 \int_0^{\sigma_1}d\tau_1 \int_0^{\tau_2} d\sigma_2 {(p_1p_2) \over
   \big(-2(p_1p_2)\sigma_1\sigma_2-2p_1Q_2\sigma_1-2p_2Q_2\sigma_2-Q_2^2 \big)^{1-\epsuv}}
    \nonumber \\
 {(p_1p_2) \over
   \big(-2(p_1p_2)\tau_1\tau_2-2p_1Q_2\tau_1-2p_2Q_2\tau_2-Q_2^2\big)^{1-\epsuv}}\ .
\end{align}

\section{Y diagram}

This diagram consists of three gluon propagators, meeting at a vertex, with two propagators ending on  side $p_1$ and the third ending on $p_2$ with sides $Q_1$ and $Q_2$ between.
The contribution of this diagram is
\begin{align}\label{y}
  {p_1 \cdot p_2\over 8 } \int_0^1 d\tau_1 d\tau_2 \Big[  2 G\big(z_1(\tau_1),z_1(\tau_1),z_2(\tau_2)\big)
  - G\big(z_1(0),z_1(\tau_1),z_2(\tau_2)\big)-G\big(z_1(1),z_1(\tau_1),z_2(\tau_2)\big)
\Big]
\end{align}
where $z_1(\tau_1)=p_1 \tau_1$ is a point ending on the edge $p_1$ and
$z_2(\tau_2)=-Q_1-p_2\tau_2$ a point on edge $p_2$ and
\begin{align}
  G(z_1,z_2,z_3)&={\Gamma(1-2\epsuv) \over \Gamma^2(1-\epsuv)}
  \int_0^1 d\alpha_1d\alpha_2d\alpha_3
  {(\alpha_1\alpha_2\alpha_3)^{-\epsuv}
    \delta(1-\alpha_1-\alpha_2-\alpha_3) \over
    (-\alpha_1\alpha_2z_{12}-\alpha_1\alpha_3z_{13}-\alpha_2\alpha_3z_{23})^{1-2\epsuv}}
\\
&={1\over \epsuv} {\Gamma(1-2\epsuv) \over \Gamma^2(1-\epsuv)} \int_0^1 d\sigma
{\sigma^{-\epsuv}(1-\sigma)^{-\epsuv} \over (-\sigma z_{13}^2 -
  (1-\sigma) z_{23}^2)^{1-2\epsuv}}
\end{align}
where the final equality is valid whenever $z_{12}^2=0$ (as is the case
here) and can be shown by changing variables to $\rho,\sigma$ with
$\alpha_1=(1-\rho) \sigma,\ \alpha_2=(1-\rho)(1- \sigma),\ \alpha_3=\rho$.

Now it turns out that the first term in \eqref{y} is precisely canceled
by half of the self-energy correction to the propagator between sides $p_1$ and
$p_2$ with the other half canceling the upside-down Y diagram.
We also notice that it is the combination of these two contributions which has the expected
maximal transcendentality.
The explicit expression for the self-energy correction to the gluon propagator
in $\cN=4$ SYM can be found, for example, in \cite{Erickson:2000af}.

Thus, neglecting the first term, the result of the Y diagram
integral is
\begin{align}
   &  {p_1\cdot p_2\over 8 } {1\over \epsuv}
    {\Gamma(1-2\epsuv) \over \Gamma^2(1-\epsuv)} \int_0^1\!\!d\sigma
    \int_0^1\!\!d\tau_1  d\tau_2
    \Big(-{\sigma^{-\epsuv}(1-\sigma)^{-\epsuv} \over (-\sigma (Q_1+p_2
    \tau_2)^2 -
  (1-\sigma) (Q_1+p_1 \tau_1+p_2
    \tau_2)^2)^{1-2\epsuv}}
     \nonumber \\
 &\qquad \qquad
  -{\sigma^{-\epsuv}(1-\sigma)^{-\epsuv} \over (-\sigma (-p_2(1-\tau_2)-Q_2)^2 -
  (1-\sigma) (Q_1+p_1 \tau_1+p_2
    \tau_2)^2 )^{1-2\epsuv}}\Big)
  \nonumber   \\[15pt]
=  &  {p_1 \cdot p_2\over 8 } {1\over \epsuv}
    {\Gamma(1-2\epsuv) \over \Gamma^2(1-\epsuv)} \int_0^1\!\!d\sigma
    \int_0^1\!\!d\tau_1  d\tau_2
    \Big(-{\sigma^{-\epsuv}(1-\sigma)^{-\epsuv} \over (-(1-\sigma) (Q_1+p_2
    \tau_2)^2 -
  \sigma (Q_1+p_1 \tau_1+p_2
    \tau_2)^2)^{1-2\epsuv}}
    \nonumber \\
 &\qquad \qquad
  -{\sigma^{-\epsuv}(1-\sigma)^{-\epsuv} \over (-(1-\sigma) (p_2 \tau_2+Q_2)^2 -
  \sigma (Q_2+p_1 \tau_1+p_2
    \tau_2)^2 )^{1-2\epsuv}}\Big)
  \nonumber   \\[15pt]
=&  {p_1\cdot p_2\over 8 } {1\over \epsuv}
    {\Gamma(1-2\epsuv) \over \Gamma^2(1-\epsuv)} \int_0^1\!\!d\sigma
    \int_0^1\!\!d\tau_1  d\tau_2
    \Big(-{\sigma^{-\epsuv}(1-\sigma)^{-\epsuv} \over (-
      Q_1^2-2 (Q_1p_2)
    \tau_2 -
  \sigma \tau_1 (2(Q_1p_1)+2(p_1p_2)\tau_2))^{1-2\epsuv}}
  \nonumber \\
 &\qquad \qquad
  -{\sigma^{-\epsuv}(1-\sigma)^{-\epsuv} \over (-(Q_2^2+2(Q_2p_2)\tau_2+
  \sigma \tau_1 (2(Q_2p_1) +2(p_1p_2)
  \tau_2) )^{1-2\epsuv}}\Big) \ ,
\end{align}
where to obtain the second equality we have used the change of
variables $\sigma \rightarrow 1-\sigma$,  $\tau_i \rightarrow
1-\tau_i$. The final answer is manifestly symmetric under $Q_1
\leftrightarrow Q_2$.

All two-loop diagrams are given by the above integrals, for various
values of the momenta $p_i$, $Q_i$.

\end{document}